%% file: Yangian-symmetry_ver_2.tex
\documentclass[11pt]{article}
\pdfoutput=1

\usepackage{jheppub}
\usepackage[lofdepth,lotdepth]{subfig} 

\usepackage{import}

\usepackage{bigints}

\title{Yangian Symmetry for the Tree Amplituhedron}
\author[1]{Livia Ferro,}\emailAdd{livia.ferro@lmu.de}
\author[2]{Tomasz \L ukowski,}\emailAdd{lukowski@maths.ox.ac.uk}
\author[1]{Andrea Orta}\emailAdd{andrea.orta@lmu.de}
\author[3]{and Matteo Parisi}\emailAdd{m.parisi@qmul.ac.uk}

\affiliation[1]{Arnold--Sommerfeld--Center for Theoretical Physics,\\ Ludwig--Maximilians--Universit\"at, \\ Theresienstra\ss e 37, 80333 M\"unchen, Germany }
\affiliation[2]{Mathematical Institute, University of Oxford,\\ Andrew Wiles Building, Radcliffe Observatory Quarter,\\ Woodstock Road, Oxford, OX2 6GG, U.K.}
\affiliation[3]{Center for Research in String Theory, \\ School of Physics and Astronomy, Queen Mary University of London \\ 327 Mile End Road, London E1 4NS, U.K.}

\abstract{Tree-level scattering amplitudes in planar $\mathcal{N}=4$  super Yang-Mills are known to be Yangian-invariant. It has been shown that integrability allows to obtain a general, explicit method to find such invariants. The uplifting of this result to the amplituhedron construction has been an important open problem. In this paper, with the help of methods proper to integrable theories, we successfully fill this gap and clarify the meaning of Yangian invariance for the tree-level amplituhedron. In particular, we construct amplituhedron volume forms from an underlying spin chain. As a by-product of this construction, we also propose a new on-shell diagrammatics for the amplituhedron. 
}

\begin{document}
\begin{flushright}
{\small LMU-ASC 63/16}\\
{\small QMUL-PH-16-21}
\end{flushright}
\maketitle

\addtocontents{toc}{\protect\setcounter{tocdepth}{1}}

\section{Introduction}
\label{introduction}

In recent years we have seen tremendous progress in developing novel, powerful formulations for scattering amplitudes. Most of the successes have been achieved  in the domain of \mbox{$\mathcal{N}=4$} super Yang-Mills theory (SYM) -- the maximally supersymmetric gauge theory in four dimensions. This model has remarkable properties, among which a predominant role is played by the high amount of symmetry, especially in the planar limit. Indeed, the standard and dual superconformal symmetries of tree-level amplitudes close into a Yangian symmetry \cite{Drummond:2009fd}.  This is governed by an infinite-dimensional Yangian algebra appearing also in integrable spin chains: an indication of the underlying quantum integrable structure of planar $\mathcal{N}=4$ SYM. 
In particular, it was shown  \cite{Drummond:2010qh,Drummond:2010uq} that the Grassmannian formula for $n$-particle N$^{k}$MHV tree-level amplitudes \cite{ArkaniHamed:2009dn,Mason:2009qx}, which in terms of momentum supertwistors \cite{Hodges:2009hk}  $\mathcal{Z}^{\mathcal{A}}_i=(\lambda_i^\alpha,\tilde\mu^{\dot \alpha}_i,\chi_i^A)$ reads
\begin{equation} \label{grass.integral}
\mathcal{A}_{n,k} = \frac1{\text{GL}(k)} \int_{\gamma} \frac{d^{k \cdot n}\,c_{\alpha i}}{(1 2\ldots k) (2 3\ldots k+1)\ldots(n 1 \ldots k-1)} \prod_{\alpha=1}^k \delta^{4|4}\left( \sum_i c_{\alpha i}  \mathcal{Z}_i\right),
\end{equation}
is  invariant under the  generators of the Yangian $Y\big(\mathfrak{psl}(4|4)\big)$
\begin{align}\label{yangiangenampls}
(J^{(0)})^\mathcal{A}_{\;\mathcal{B}}&=\sum_{i=1}^n \mathcal{Z}_i^\mathcal{A}\frac{\partial}{\partial \mathcal{Z}^\mathcal{B}_i} \,,\nonumber \\
(J^{(1)})^\mathcal{A}_{\;\mathcal{B}}&=\sum_{i<j} \left(\mathcal{Z}_i^\mathcal{A}\frac{\partial}{\partial \mathcal{Z}^\mathcal{C}_i}\mathcal{Z}_j^\mathcal{C}\frac{\partial}{\partial \mathcal{Z}^\mathcal{B}_j}-(i\leftrightarrow j)\right) \,.
\end{align}
The Grassmannian formulation initiated a new way of thinking about amplitudes which led, more recently, to the amplituhedron proposal \cite{Arkani-Hamed:2013jha}. In this framework, every amplitude of planar $\mathcal{N}=4$ SYM  is conjectured to be the volume of a novel object, the (dual) amplituhedron.
The question then arises as to which symmetries the amplituhedron possesses and, in particular, whether it inherits the Yangian symmetry of tree-level amplitudes. In analogy with the representation \eqref{grass.integral} of scattering amplitudes, the present authors proposed  in a previous paper \cite{Ferro:2015grk} the following expression for the so-called volume functions:
\begin{equation} 
\label{omega}
\Omega_{n,k}^{(m)}(Y,Z) = \int_{\gamma} \frac{d^{k \cdot n}\,c_{\alpha i}}{(1 2\ldots k) (2 3\ldots k+1)\ldots(n 1 \ldots k-1)} \prod_{\alpha=1}^k \delta^{m+k}(Y_{\alpha} - \sum_i c_{\alpha i}  Z_i) \,,
\end{equation}
where the $Z_i$ are a bosonized version of the momentum supertwistors $\mathcal{Z}_i$ and $Y_{\alpha}$ are $k$ auxiliary vectors, as we will explain in the main text. As already proposed in the literature, we consider here a more general version of the problem by introducing an additional parameter $m$, related to the number of spacetime dimensions which, for scattering amplitudes in $\mathcal{N}=4$ SYM, should be set to $m=4$. In \cite{Ferro:2015grk} we showed that $\Omega_{n,k}^{(m)}(Y,Z)$ satisfies a set of differential equations, corresponding to $GL(m+k)$-covariance and scaling properties, together with the so-called Capelli differential equations. Altogether they completely  determine the volume for next-to-maximally helicity violating (NMHV) amplitudes. 
For higher-helicity cases, however,  the volume is only partially constrained and therefore it is natural to ask whether Yangian invariance, if present, can fix it completely.
The volume function $\Omega_{n,k}^{(m)}(Y,Z)$, despite being very reminiscent  of  \eqref{grass.integral}, is however not Yangian invariant, at least  under the simple ``bosonization" of the generators \eqref{yangiangenampls} which would lead to the Yangian $Y\big(\mathfrak{gl}(m+k)\big)$.\footnote{Actually, the Capelli differential equations imply Yangian symmetry  $Y\big(\mathfrak{gl}(m+1)\big)$ for NMHV amplitudes.}
Indeed, the inclusion of the auxiliary vectors $Y_{\alpha}$ and the purely bosonic description of the amplituhedron spoil the properties which were essential in showing the Yangian invariance of \eqref{grass.integral}. %
In particular, already when acting with level-zero generators one obtains non-vanishing expressions, which would however trivially integrate to zero when extracting the amplitude from the volume form. This directly follows from the fact that monomials of sufficiently high order in a finite number of Grassmann-odd variables evaluate to zero. In this paper we show that, although the generators of $Y\big(\mathfrak{gl}(m+k)\big)$ do not annihilate the volume function, the expressions we get belong to the kernel of a simple differential operator.  As a result, we will prove that there exists a matrix of functions closely related to the amplituhedron volume function which is invariant under the Yangian of $\mathfrak{gl}(m+k)$. 
To this purpose, we follow the steps of \cite{Kanning:2014maa}, where Yangian invariants relevant for tree-level amplitudes in  $\mathcal{N}=4$ SYM have been obtained using the Quantum Inverse Scattering Method.
Indeed, the infinite-dimensional symmetry algebra governing planar $\mathcal{N}=4$ SYM allows us to employ methods and techniques proper to (or inspired by) integrable theories. The first step in this direction was taken in \cite{Ferro:2013dga,Ferro:2012xw}: there, a deformation of scattering amplitudes in terms of a spectral parameter was proposed by considering Yangian generators relevant to inhomogeneous spin chains. 
This has led, in particular, to a systematic approach for the construction of Yangian invariants relevant for scattering amplitudes from an underlying spin chain description  \cite{Kanning:2014maa,Frassek:2013xza,Beisert:2014qba,Broedel:2014pia}. This story generalizes to the amplituhedron, as we show in this paper.

The paper is organized as follows.
In section \ref{basics}, we review some notions related to the tree-level amplituhedron. 
In section \ref{onshelldiagrs} we adapt the on-shell diagrammatics for amplitudes and introduce its avatar, relevant for volume functions. In section \ref{spinchain} we show how it follows directly from a spin chain construction. 
Finally, in section \ref{invariance}, we discuss the Yangian invariance of $\Omega_{n,k}^{(m)}(Y,Z)$ and relate it to diffeomorphisms of the positive Grassmannian.
We end with conclusions and outlook.
More technical results are postponed to the appendices.


\section{The Tree-Level Amplituhedron and its Volume}
\label{basics}

To define the tree-level amplituhedron one introduces a bosonized version of the momentum supertwistors $\mathcal{Z}^{\mathcal{A}}_i$ \cite{ArkaniHamed:2010gg}: one defines new variables $Z^{A}_i$, whose components include the bosonic part of the  supertwistors $z_i=(\lambda_i^\alpha,\tilde\mu^{\dot \alpha}_i)$, supplemented by a bosonized version of the fermionic components $\xi^a_{i}=\phi^a_{\mathsf{A}}\chi^\mathsf{A}_i$, $a=1,\dots,k$. Here the $\phi^a_{\mathsf{A}}$ are auxiliary Grassmann-odd parameters and $k$ labels the helicity sector of the superamplitude. The variables $Z^{A}_i$ will be called {\it bosonized momentum twistors}. In the physical setting we have $A=1,\ldots ,4+k$, however we will allow this index to range over $1,\ldots, m+k$, for any even value of $m$. In particular, the case $m=2$ is often a good testing ground for our ideas.  
	
Let us demand that the bosonized twistors be positive, $Z=(Z^A_i)\in M_+(m+k,n)$, where $M_+(m+k,n)$ is the set of $( m+k) \times n$ positive real matrices, {\it i.e.}~matrices whose ordered maximal minors are positive:
\begin{equation}\label{positivity}
\langle Z_{i_1}\dots Z_{i_{m+k}} \rangle > 0 \qquad,\qquad \textrm{for}\qquad
1\leq i_1<\ldots<i_{m+k}\leq n\,.
\end{equation}
The tree-level amplituhedron is defined as the space \cite{Arkani-Hamed:2013jha}
\begin{equation}\label{defamplituhedron}
\mathtt{A}^{\mathrm{tree}}_{n,k;m}[Z] := \bigg\{ Y=(Y^A_\alpha) \in G(k,m+k) \;\;:\;\; Y^A_{\alpha}=\sum_i c_{\alpha i} Z^A_i \, ,\quad C=(c_{\alpha i}) \in G_+(k,n) \bigg\} \, ,
\end{equation}
where $G_+(k,n)$ is the positive Grassmannian, {\it i.e.}~the restriction of $G(k,n)$ to the matrices with positive ordered $k \times k$ minors.
One can canonically define a $(k\cdot m)$-dimensional differential form $\mathbf{\Omega}_{n,k}^{(m)}(Y,Z)$ on $\mathtt{A}^{\mathrm{tree}}_{n,k;m}[Z] $ demanding that it has logarithmic singularities on all boundaries of the space \cite{Arkani-Hamed:2013jha}: in terms of local coordinates, this means that it must behave as $d \alpha / \alpha$ when approaching any boundary. Such an object is called {\it volume form} and in general it can be written as
\begin{equation}\label{volume.form}
\mathbf\Omega_{n,k}^{(m)}(Y,Z)=\prod_{\alpha=1}^k \langle Y_1\ldots Y_k d^{m} Y_\alpha \rangle \, \Omega_{n,k}^{(m)}(Y,Z)\,,
\end{equation}
where we introduced the {\it volume function} $\Omega_{n,k}^{(m)}(Y,Z)$, the main object of interest in the following.
As already pointed out in the Introduction,  volume functions admit the following integral representation  \cite{Ferro:2015grk}:
\begin{equation} \label{omegaintegral}
\Omega_{n,k}^{(m)}(Y,Z) = \int_\gamma \frac{d^{k \cdot n}\,c_{\alpha i}}{(1 2\ldots k) (2 3\ldots k+1)\ldots(n 1 \ldots k-1)} \prod_{\alpha=1}^k \delta^{m+k}(Y_{\alpha}^A - \sum_i c_{\alpha i}  Z_i^A)\,,
\end{equation}
where the integral is evaluated along a closed contour $\gamma$. Similarly to what was presented in \cite{Arkani-Hamed:2013jha}, one can extract the Grassmannian integral \eqref{grass.integral} from $\Omega_{n,k}^{(4)}(Y,Z)$ by localizing $Y$ on some reference point, {\it e.g.}~$Y^*=(0_{m\times k} \big| \mathbb{I}_k)^T$, and integrating over the auxiliary fermionic parameters $\phi^a_{ \mathsf{A}}$, namely
\begin{equation}\label{from.volume.to.amplitude}
\mathcal{A}_{n,k}=\int d^{4 k}\phi^{a}_{\mathsf{A}} \; \Omega^{(4)}_{n,k}(Y^*,Z)\,.
\end{equation}
 Then, the $n$-particle N$^k$MHV amplitude is determined by a proper choice of contour $\gamma$, which can be fixed {\it e.g.}~by the BCFW recursion and allows to compute the scattering amplitude as a particular combination of residues of \eqref{grass.integral}. Each such residue can be associated to a particular cell of the positive Grassmannian $G_+(k,n)$, with the original integral \eqref{grass.integral} corresponding to the so-called {\it top cell} of $G_+(k,n)$, see \cite{ArkaniHamed:2012nw,*ArkaniHamed:2012nwB}. 
 
Let us stress once more that the main obstacle to the direct, naive translation of supersymmetric Yangian generators into the amplituhedron language is the fermionic integration in \eqref{from.volume.to.amplitude}. The significance of the construction presented in this paper is related to the fact that, due to the Grassmann-odd variables hidden in the  $\xi_i^a$, non-vanishing expressions arise which integrate to zero at the amplitude level. It is therefore a non-trivial task to find generators annihilating the volume function itself, before the integration is carried out.


\section{Construction of Amplituhedron Volume Functions}
\label{onshelldiagrs}

Before discussing the Yangian invariance of amplituhedron volume functions $\Omega_{n,k}^{(m)}(Y,Z)$, we show in this section how to construct \eqref{omegaintegral} by introducing an on-shell diagrammatics similar to, but different from, the one established for scattering amplitudes. In the following section we argue that this diagrammatics follows directly from an underlying spin chain description for the tree-level amplituhedron. Both derivations will parallel a similar construction for \eqref{grass.integral} explained in detail in \cite{Kanning:2014maa}. The main ingredient there is the operator \cite{Chicherin:2013ora}
\begin{equation}\label{Boperator}
\mathcal{B}_{ij}(u)=\left(\mathcal{Z}_j^\mathcal{A}\partial_{\mathcal{Z}_i^\mathcal{A}}\right)^u=\mathcal{N}\int \frac{d\alpha}{\alpha^{1+u}}e^{\alpha\mathcal{Z}_j^\mathcal{A}\partial_{\mathcal{Z}_i^\mathcal{A}}} \,,
\end{equation}
where the normalization $\mathcal{N}$ and the integration contour will not be relevant to our discussion. In the following we will also be using the bosonized version of \eqref{Boperator}: it will be clear from the context which definition we will be working with. Notice that the rightmost expression in \eqref{Boperator} is well defined even when $u=0$. 

\subsection{From Scattering Amplitudes to the Amplituhedron}

Let us start by reviewing the main steps in the construction of \eqref{grass.integral}. One can associate this Grassmannian integral with the top cell of the positive Grassmannian $G_+(k,n)$. As explained in \cite{Postnikov:2006kva}, to each cell of its positroid stratification one can in turn associate a permutation of the symmetric group $S_n$. For the top cell of $G_{+}(k,n)$ this takes the particularly simple form
\begin{equation}\label{topcellsigma}
\sigma_{n,k}(i)=i+k \,(\mathrm{mod}\, n)\,.
\end{equation}
Next, we need to decompose $\sigma_{n,k}$ into adjacent transpositions. Let us emphasize that such a decomposition is not unique but the function which we will associate to the given permutation will be independent of it. Moreover, we assume to be working with decompositions with the least number of factors, called \emph{minimal} decompositions.  One example  we will use throughout this paper is 
\begin{equation}\label{topcelldecomposition}
\sigma_{n,k}=\underbrace{(k,k+1)\ldots(n-1,n)}_{n-k \text{ factors}}\ldots\underbrace{(23)\ldots(n-k+1,n-k+2)}_{n-k \text{ factors}}\underbrace{(12)\ldots(n-k,n-k+1)}_{n-k \text{ factors}}	\,,
\end{equation}
where $(i,j)$ denotes the permutation swapping $i$ and $j$. We remark that the composition of transpositions in \eqref{topcelldecomposition} has to be understood as acting on the identity by swapping preimages (images) when reading from the left (right).
It is easy to find an explicit form for the $l$-th factor $(i_l, j_l)$ of the above decomposition (counting from the right): if we write $l = p(n-k) + q$, where $q \in \{1,2,\dots,n-k\}$, from the explicit form of $\sigma_{n,k}$ we get
\begin{equation}\label{subscript.formula}
i_l = n - k + p - q + 1 \, ,\quad j_l = i_l + 1 \,.
\end{equation}
Provided the decomposition \eqref{topcelldecomposition}, one can show that \eqref{grass.integral} can be constructed as
\begin{equation}\label{amplitudsc}
\mathcal{A}_{n,k}(\mathcal Z) =\prod_{l=1}^{k(n-k)}\mathcal{B}_{i_l j_l}(0) \prod_{i=1}^k \delta^{4|4}(\mathcal{Z}^\mathcal{A}_i) \,,
\end{equation}
where the operators $\mathcal B_{i_l j_l}$ appear in the opposite order compared to the order of factors of $\sigma_{n,k}$. This construction is based on the possibility of building up any cell of the positive Grassmannian $G_+(k,n)$ -- and the associated canonical forms -- starting from zero-dimensional cells, which correspond to vacua of the spin chain.

A crucial difference between the integral for volume functions \eqref{omegaintegral} and its analogue for scattering amplitudes \eqref{grass.integral} is the presence of the auxiliary variables $Y_\alpha^A$. It turns out that this difference can be traced back to the choice of ``vacuum'' on which the operators $\mathcal{B}_{ij}$ act. For scattering amplitudes it is a product of $\delta$-functions which,  in the amplituhedron context, has to be replaced by the \emph{seed} $\mathcal{S}_k^{(m)}$
\begin{equation}
\prod_{i=1}^k \delta^{m|m}(\mathcal{Z}^\mathcal{A}_i)  \longrightarrow \mathcal{S}_k^{(m)} \,,
\end{equation}
defined as 
\begin{equation}\label{seed.vertex}
\mathcal{S}_k^{(m)} := \bigintssss \frac{d^{k\cdot k}\beta}{(\det \beta)^k}\, \prod_{\alpha=1}^k \delta^{m+k}\bigg(Y^A_\alpha-\sum_{i=1}^k\beta_{\alpha i} Z_i^A \bigg)	\,.
\end{equation}
Notice that $\mathcal{S}_k^{(m)}$ involves only $k$ bosonized momentum twistors, which we have chosen to be $Z_1,\dots, Z_k$. The result is independent of this choice, provided that the $Z_i$ are consecutive, which guarantees cyclic invariance of the volume form. In the following we show that the cyclicity relies on a new set of transformations, which extends the usual set of cluster mutations present for the on-shell diagrams relevant to scattering amplitudes.
It is then straightforward to check that the volume function can be written as 
\begin{equation}\label{omega.from.seed}
\Omega_{n,k}^{(m)}(Y,Z) = \prod_{l=1}^{k(n-k)}\mathcal{B}_{i_l j_l}(0)\,\mathcal{S}_k^{(m)} \,,
\end{equation}
see Appendix \ref{app.omega.proof} for details. This is the main formula of this section and we will use it in the following to establish a connection between volume functions and spin chains.

Our discussion focused so far on the top cell of the positive Grassmannian. However, there is a natural way to generalize it to any residue of \eqref{omegaintegral}. As we mentioned already, all its residues are in  one-to-one correspondence with the cells $C_\sigma$ of the positive Grassmannian $G_+(k,n)$, which in turn are labelled by the permutations $\sigma \in S_n$. In order to find a formula similar to \eqref{omega.from.seed} for a given residue, which we denote $\Omega^{(m)}_\sigma(Y,Z)$, we first decompose the permutation $\sigma$ into generalized adjacent transpositions
\begin{equation}
\sigma = \prod_{l=1}^{|\sigma|}(\mathsf i_l, \mathsf j_l) = (\mathsf i_{|\sigma|}, \mathsf j_{|\sigma|}) \ldots (\mathsf i_2, \mathsf j_2) (\mathsf i_1, \mathsf j_1) \,,
\end{equation}
where $|\sigma|$ is the dimension of the cell $C_\sigma$. A decomposition into generalized adjacent transpositions is characterized by the following condition: any $\mathsf j_l$ is allowed to be bigger than $\mathsf i_l+1$, provided that for every $q > p$ one has $\mathsf i_q, \mathsf j_q \notin \{\mathsf i_p+1, \ldots, \mathsf j_p -1\}$.  Then one can show that 
\begin{equation}\label{omega.sigma}
\Omega_\sigma^{(m)}(Y,Z) = \prod_{l=1}^{|\sigma|}\mathcal{B}_{\mathsf i_l \mathsf j_l}(0)\,\mathcal{S}_k^{(m)} \,.
\end{equation}
This gives us a very concrete prescription to calculate the volume functions associated to the individual BCFW terms contributing to a given amplitude. The permutations labelling them -- which can be computed by means of the program \texttt{positroids} \cite{Bourjaily:2012gy} -- are naturally given in ordinary twistor language: after translating them into momentum twistor formulation \cite{ArkaniHamed:2012nw,*ArkaniHamed:2012nwB}, their decomposition into adjacent transpositions provides us with the labels $(\mathsf i_l, \mathsf j_l)$ needed in \eqref{omega.sigma}. Finally, the sum of the resulting $\Omega_\sigma^{(m)}(Y,Z)$ is the sought-after volume function.

\subsection{On-shell Diagrammatics}

\begin{figure}[h!]
\centering
\begin{tabular}{ccccl}
\def\svgwidth{2cm}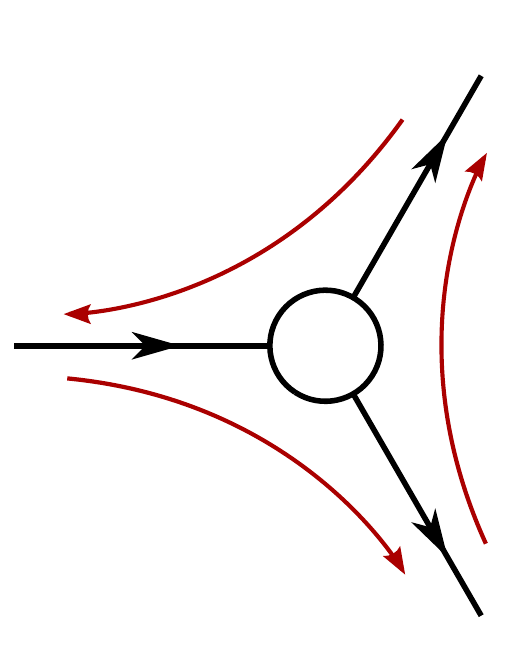 && \raisebox{1.2cm}{$\longrightarrow$} &&
\raisebox{1.2cm}{$\displaystyle \bigintssss \frac{d c_2 d c_3}{c_2 c_3}\, \delta^{(m+k)}\big(Z_1^A+c_2 Z_2^A+c_3 Z_3^A \big)$} \\
\def\svgwidth{2cm}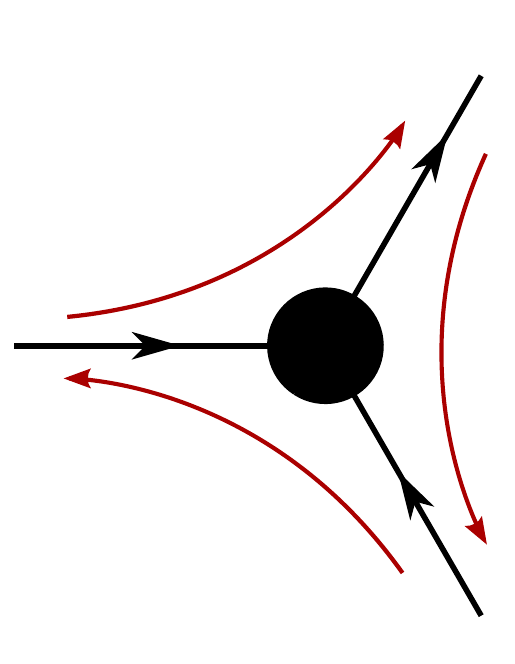 && \raisebox{1.2cm}{$\longrightarrow$} &&
\raisebox{1.2cm}{$\displaystyle\bigintssss \frac{d c_1 d c_2}{c_1 c_2}\, \delta^{(m+k)}\big(Z_1^A+c_1 Z_3^A \big)\delta^{(m+k)}\big(Z_2^A+c_2 Z_3^A \big)$}
\end{tabular}
\caption{Trivalent vertices.}\label{Fig.(anti)MHV3}
\end{figure}

\begin{figure}[h!]
\centering
\begin{tabular}{lcccl}
\hspace{-1.25cm} \def\svgwidth{2cm}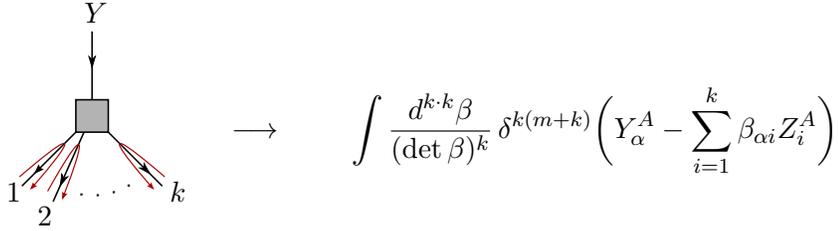 \hspace{-0.25cm} && \hspace{0cm} \raisebox{1.2cm}{$\longrightarrow$}\hspace{0cm}  &&
\hspace{-0cm} \raisebox{1.2cm}{$\displaystyle \bigintssss \frac{d^{k \cdot k}\beta}{(\det \beta)^k}\, \delta^{k(m+k)}\bigg(Y^A_\alpha-\sum_{i=1}^k\beta_{\alpha i}  Z_i^A \bigg)$}
\end{tabular}
\caption{Seed vertex, corresponding to the function $\mathcal S_k^{(m)}$.}\label{Fig.SeedVertex}
\end{figure}

The above discussion suggests an on-shell diagrammatics for volume functions bearing several similarities to that relevant to scattering amplitudes. There, one could construct all Yangian invariants using just two vertices corresponding to the MHV and the $\overline{\mathrm{MHV}}$ three-point amplitudes \cite{ArkaniHamed:2012nw,*ArkaniHamed:2012nwB}; in the case of the amplituhedron, those vertices are modified and their explicit form can be found in Fig.~\ref{Fig.(anti)MHV3}. Notice that, opposed to the amplitude case, the parameter $k$ associated to the full diagram appears explicitly at each vertex via the $\delta$-functions. Moreover, the arrows on the edges of the diagrams indicate the gauge-fixing we use: we can evaluate only those diagrams which can be given a \emph{perfect orientation}, \textit{i.e.} such that all of their trivalent vertices can be dressed with arrows as depicted below.  
We also need to introduce the new seed vertex $\mathcal{S}_k^{(m)}$ corresponding to the vacuum, depicted in Figure \ref{Fig.SeedVertex}. These three ingredients are enough to graphically represent the residue \eqref{omega.sigma} associated to any cell of the positive Grassmannian. Let us mention that, interestingly, a similar new type of seed vertex (for $k=2$) was introduced in the context of on-shell diagrams for form factors \cite{Frassek:2015rka}. The permutation labelling a given invariant or, equivalently, a cell of the positive Grassmannian, can be read from the corresponding diagram by following its edges from one external leg to another (its image), taking right (left) turns at every white (black) vertex, as shown in Figure \ref{Fig.(anti)MHV3}. Moreover, one has to turn back when encountering the seed vertex $\mathcal{S}_k^{(m)}$.


\begin{figure}[h!]
\centering
\subfloat[][$k=1$ seed cluster mutation]{\label{Fig.ClustMut1}
\def\svgwidth{5.85cm}
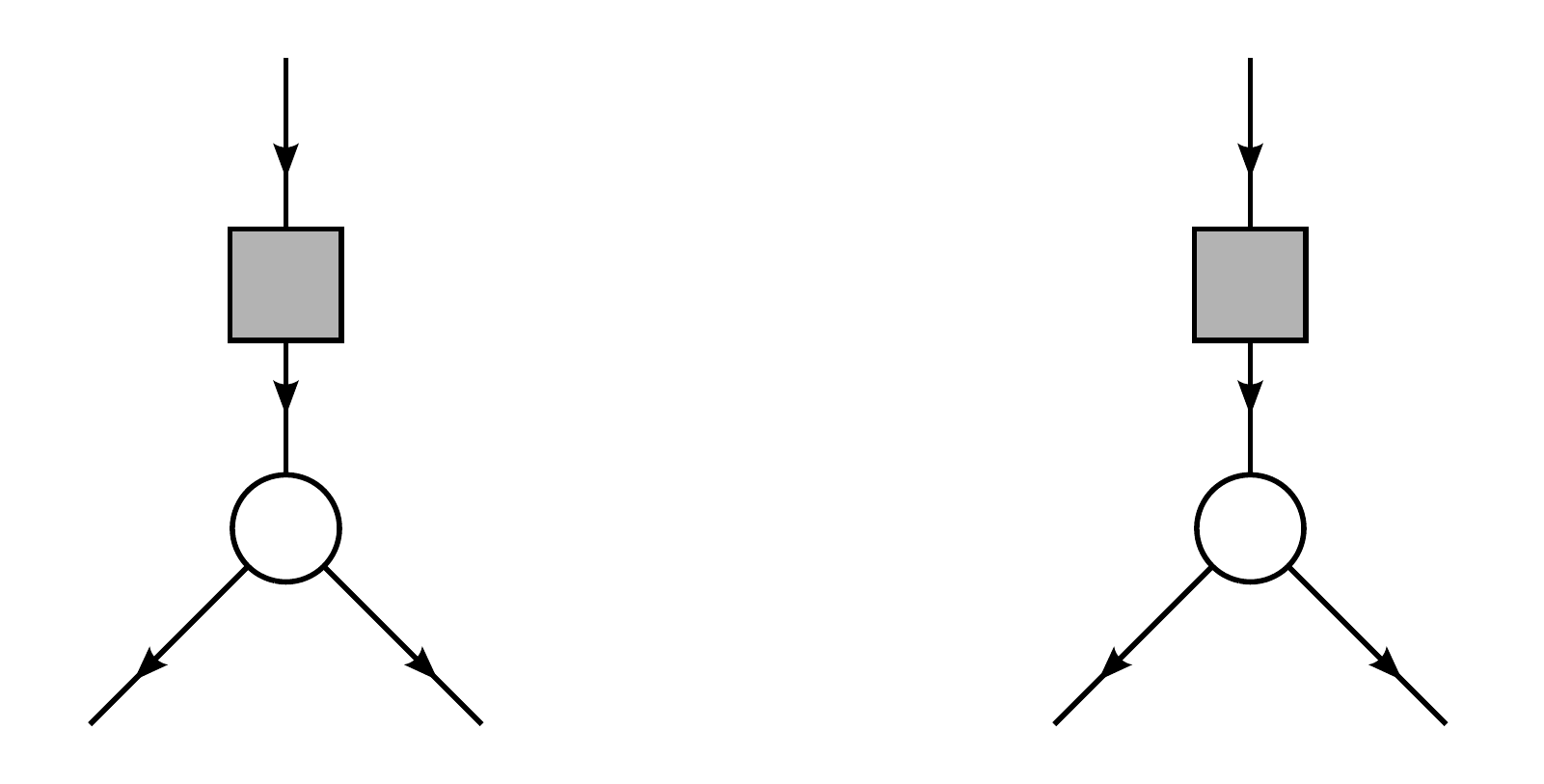} \qquad\qquad
\subfloat[][$k=2$ seed cluster mutation]{\label{Fig.ClustMut2}
\def\svgwidth{7.5cm}
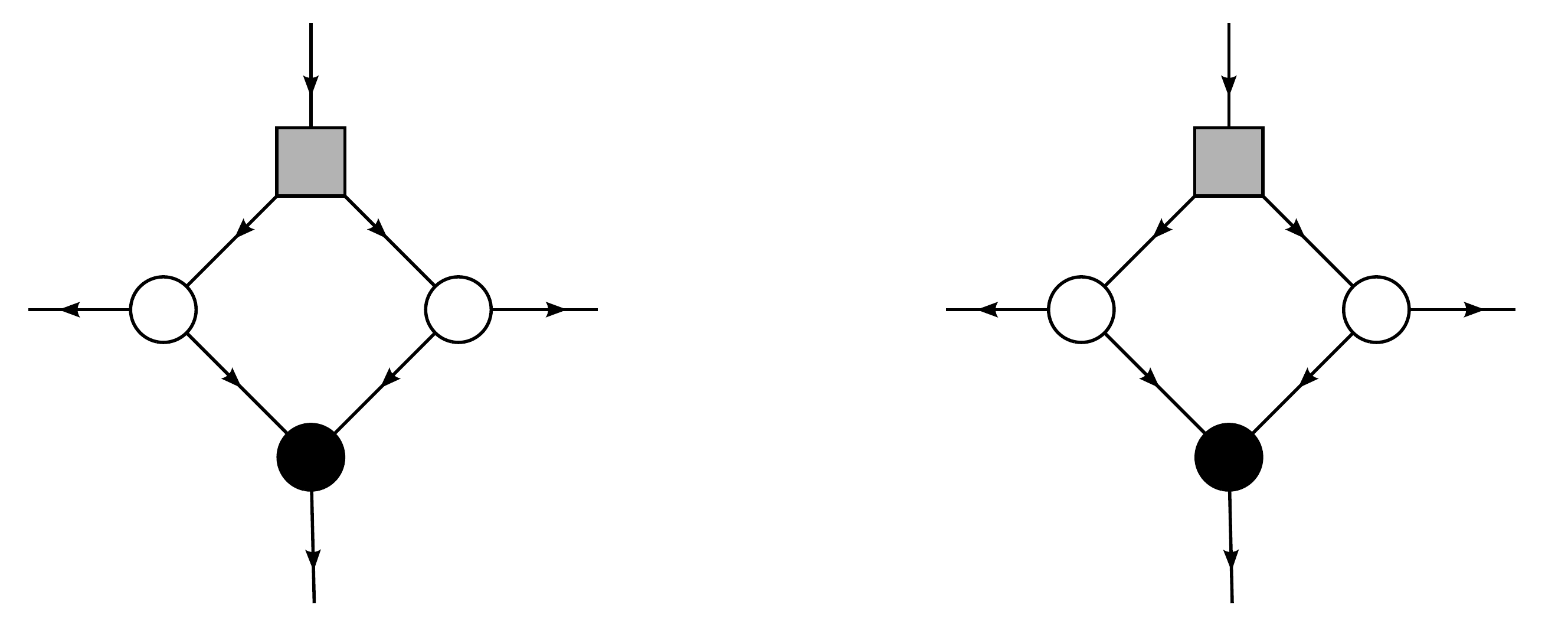} \\
\subfloat[][$k=3$ seed cluster mutation]{\label{Fig.ClustMut3}
\def\svgwidth{10cm}
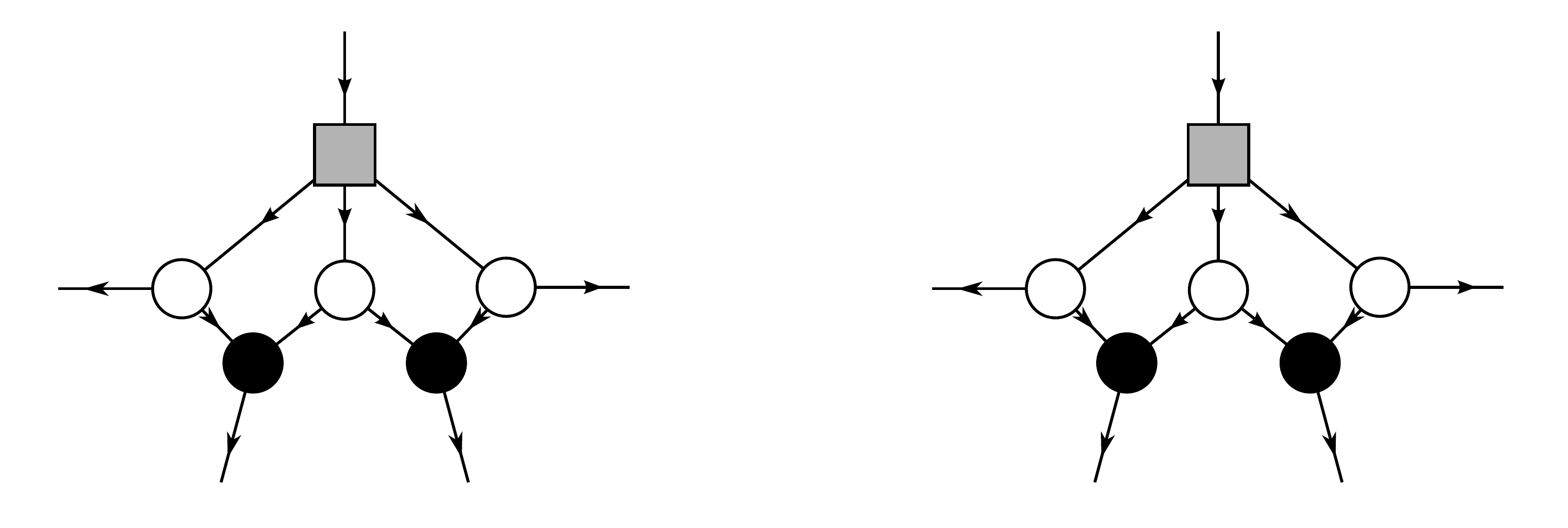} 
\caption{Cluster mutations of the seed vertex for $k=1,2,3$}
\label{Fig.ClusterMutations}
\end{figure}

Just as for scattering amplitudes, different on-shell diagrams might evaluate to the same volume function. In the former case, such equivalent graphs can be mapped to each other using transformations preserving the corresponding functions, the so-called square and flip moves. Obviously, the presence of a third kind of vertex begs the question of whether some additional cluster structure exists, which further constrains the number of actually distinct diagrams. This is indeed the case and Figure \ref{Fig.ClusterMutations} displays the transformations -- which we call \emph{seed cluster mutations} -- under which NMHV, N$^2$MHV and N$^3$MHV volume functions are invariant. It is immediate to extrapolate the pattern for arbitrary $k$ and a general proof of their validity is provided in Appendix \ref{App.seed.mutation}. 
Cluster mutations allow in particular to prove cyclic invariance of the volume functions: an appropriate sequence of flip and square moves and seed cluster mutations maps a given on-shell diagram to a version of itself where the external legs have been cyclically relabelled. 

As a further remark, it was shown in \cite{Kanning:2014maa} that it is also possible to construct deformed Grassmannian integrals following a similar procedure to that of \eqref{omega.sigma}. So far we have used solely the operators $\mathcal{B}_{ij}(0)$: in order to obtain deformed Grassmannian integrals, we need to allow a non-trivial dependence on $u$-parameters. Then, demanding that the obtained integrals are Yangian-invariant, it was found in \cite{Kanning:2014maa} that they are again in one-to-one correspondence with permutations and they smoothly reduce to undeformed integrals when the deformations are removed. In the next section we will pursue an analogous approach, yielding an explicit construction of {\it deformed} volume functions.

\begin{figure}[h!]
\centering
\subfloat[][Bridge construction]{\label{Fig.n4k2_Bridges}
\def\svgwidth{4cm}
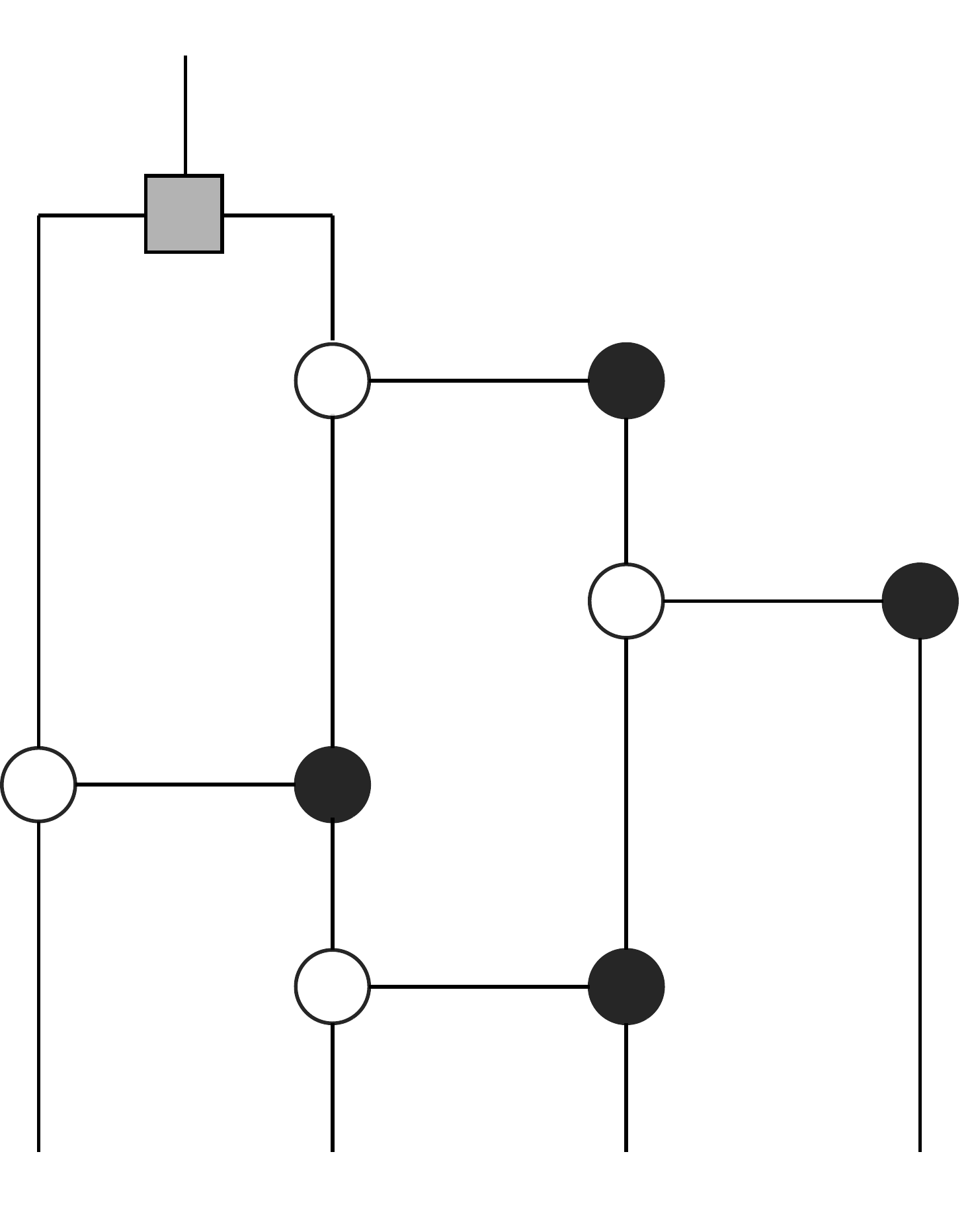} \qquad\qquad\qquad\qquad
\subfloat[][\mbox{On-shell diagram}]{\label{Fig.n4k2_OnShellDiagram}
\def\svgwidth{5cm}
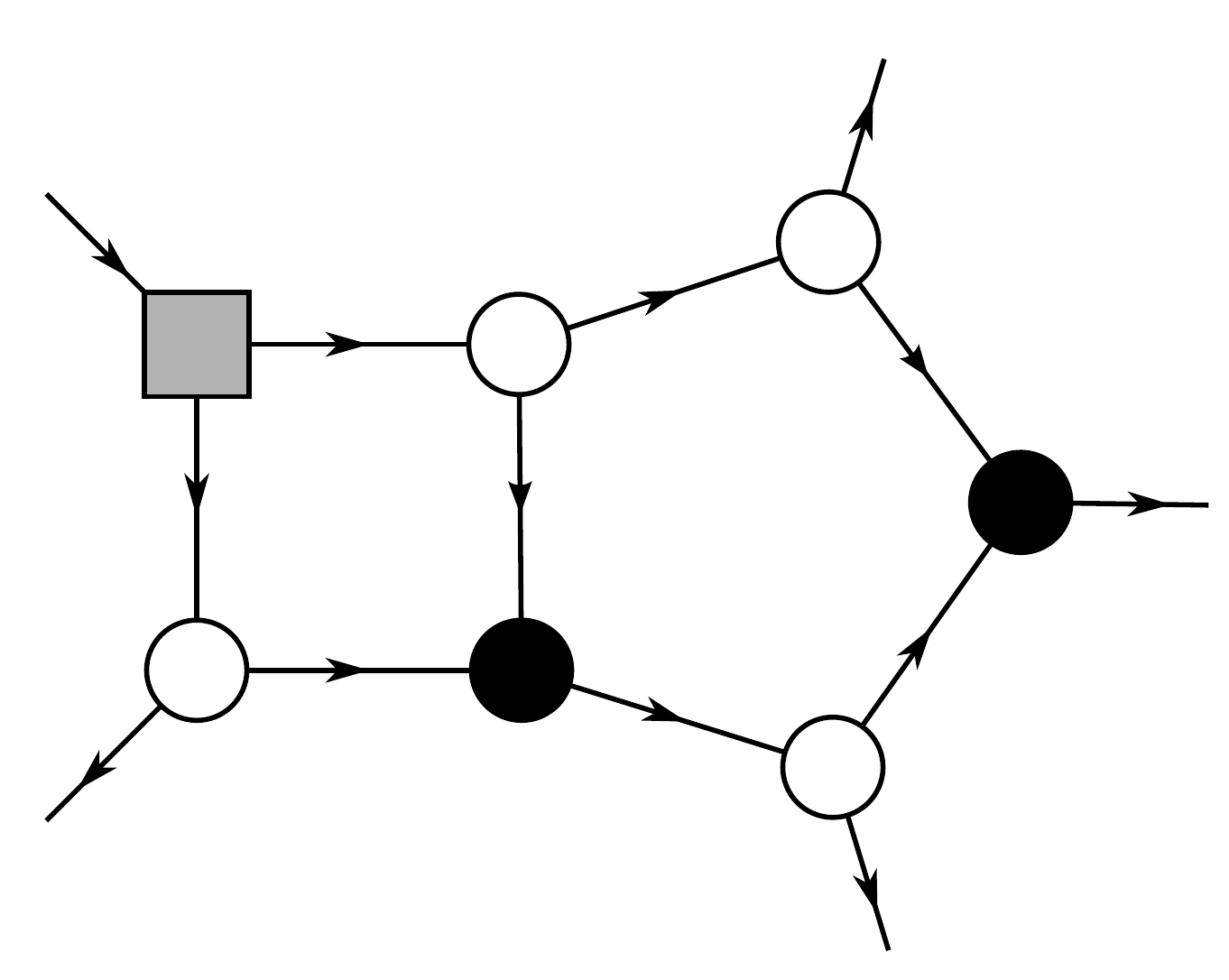} \quad\quad\quad
\caption{Diagrammatic representation for the $n=4$, $k=2$ volume function.}
\label{Fig.n4k2}
\end{figure}

We conclude this section with a simple example illustrating all concepts we have introduced so far. More examples (in the context of scattering amplitudes) can be found in \cite{Kanning:2014maa}. Let us consider the case of $\Omega_{4,2}^{(m)}(Y,Z)$. The top-cell permutation is
\begin{equation}
\sigma_{4,2} = \left( \begin{tabular}{cccc}1&2&3&4\\3&4&1&2\end{tabular}\right)=(23)(34)(12)(23)\,.
\end{equation}
Then, according to \eqref{omega.from.seed}, we have
\begin{equation}\label{omega.42}
\Omega_{4,2}^{(m)}(Y,Z)=\mathcal{B}_{23}(0)\mathcal{B}_{12}(0)\mathcal{B}_{34}(0)\mathcal{B}_{23}(0) \,\mathcal{S}_2^{(m)}\,.
\end{equation}
This procedure can be depicted as in Figure \ref{Fig.n4k2} where $\mathcal{B}_{ij}$ is represented as the so-called BCFW bridge composed of one black and one white trivalent vertex. On the right we depict the corresponding on-shell diagram for $\Omega_{4,2}^{(m)}(Y,Z)$ which can be obtained by removing all bivalent vertices. 

\noindent Computing the volume form via \eqref{omega.42}, we obtain
\begin{equation}
\Omega_{4,2}^{(m)} = \int \frac{d^4 \beta}{(\det\beta)^2}\frac{d\alpha_1 d\alpha_2 d\alpha_3 d\alpha_4}{ \alpha_1 \alpha_2 \alpha_3 \alpha_4} \; \delta^{2\cdot (2+m)} (Y - \beta\cdot \tilde C \cdot Z) \,, \quad 
\tilde C = \begin{pmatrix} 1 & \alpha_3 & \alpha_3\alpha_4 & 0 \\ 0 & 1 & \alpha_1 + \alpha_4 & \alpha_1\alpha_2
\end{pmatrix} \, ,
\end{equation}
\noindent Although for $m=4$ this function is overly constrained and therefore vanishes for generic external data, for $m=2$ one gets the well-known result
\begin{equation}\label{omega42}
\Omega_{4,2}^{(2)} = \frac{\langle1234\rangle^2}{\langle Y12\rangle\langle Y23\rangle\langle Y34\rangle\langle Y41\rangle} \, .
\end{equation}

\begin{figure}[h!]
\centering
\subfloat[][Bridge construction]{\label{Fig.n4k2_Bridges_alt}
\def\svgwidth{4cm}
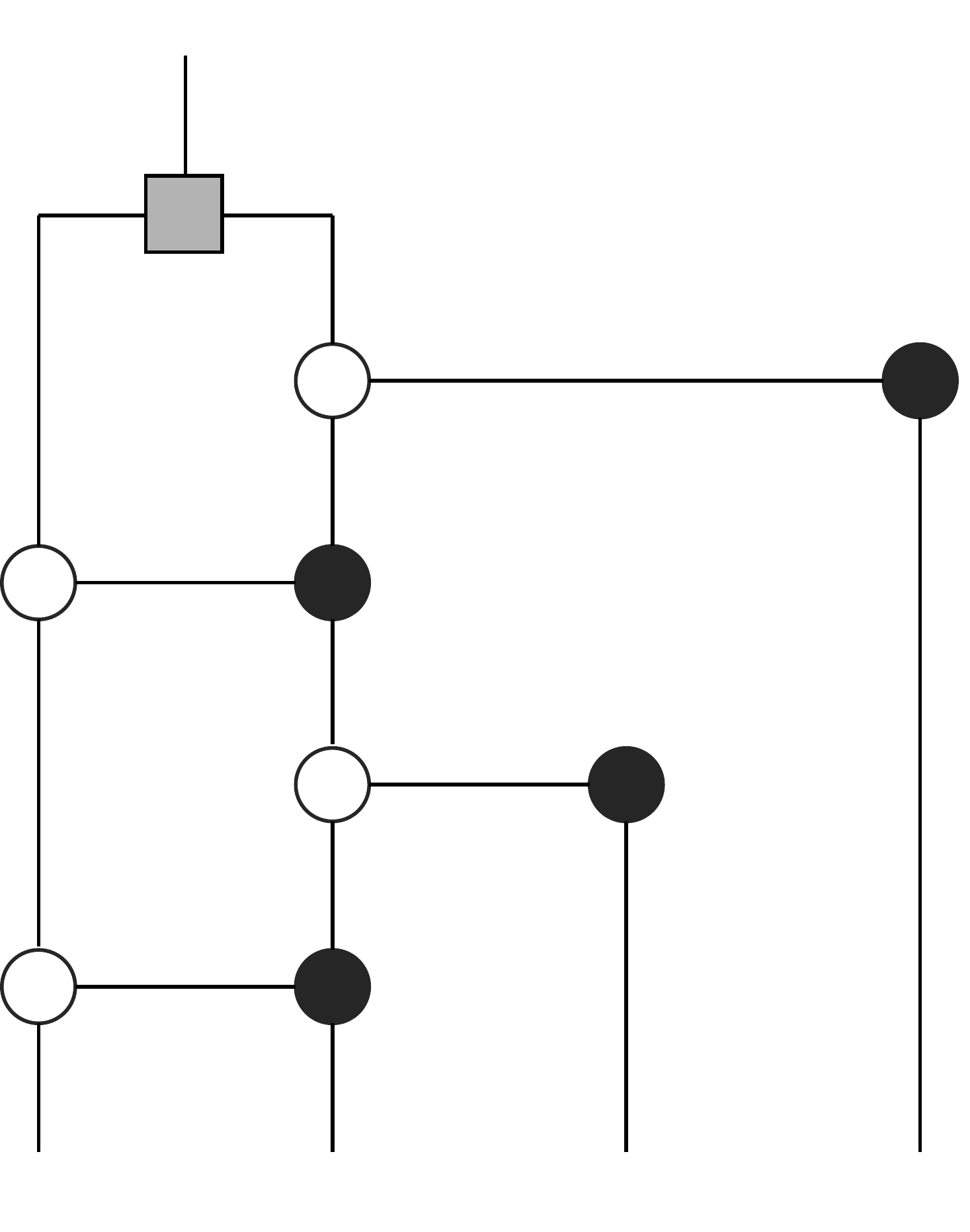} \qquad\qquad\qquad\qquad
\subfloat[][\mbox{On-shell diagram}]{\label{Fig.n4k2_OnShellDiagram_alt}
\def\svgwidth{5cm}
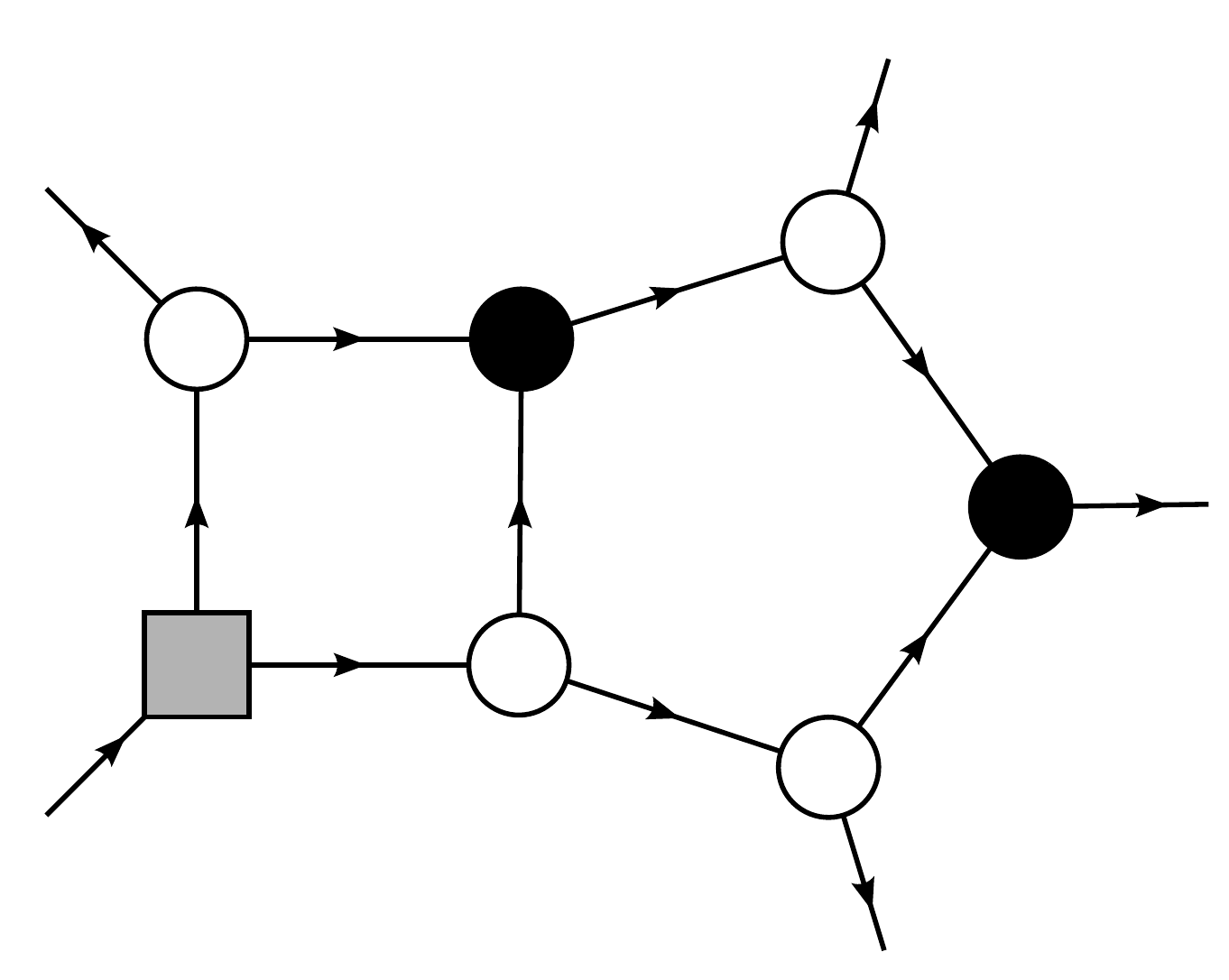} \quad\quad\quad
\caption{Another decomposition for the $n=4$, $k=2$ volume function.}
\label{Fig.another.decomposition}
\end{figure}

In order to clarify the fact that the volume function does not depend on the representation of the associated permutation in terms of transpositions, let us focus on the alternative decomposition $\sigma_{4,2} = (24)(12)(23)(12)$. The corresponding bridge construction and on-shell diagram are depicted in Fig.~\ref{Fig.another.decomposition}. It is straightforward to explicitly calculate the function associated to this on-shell diagram and find again \eqref{omega42}. This equivalence can also be shown using a simple sequence of square and flip moves. Moreover, a seed cluster mutation as in Fig.~\ref{Fig.ClustMut2} yields the diagram in Fig.~\ref{Fig.n4k2_OnShellDiagram} with cyclically relabelled external legs, proving diagrammatically the cyclic invariance of \eqref{omega42}.

\section{Spin Chain Construction of the Amplituhedron Volume}
\label{spinchain}
As we already mentioned in the introduction, the action of the  generators of $Y\big(\mathfrak{gl}(m+k)\big)$ on the volume function is not zero. In this section we will show that there exists an operator, namely
\begin{equation}
(J_Y)^A_{\; B}=\sum_{\alpha=1}^k\frac{\partial}{\partial Y^B_\alpha}Y_\alpha^A=\sum_{\alpha=1}^k Y^A_\alpha\frac{\partial}{\partial Y_\alpha^B}+k\,\delta^A_{\;B} \;,
\end{equation}
for which
\begin{equation}\label{operator.JY}
(J_Y)^A_{\;C}\,(J^{(\ell)})^C_{\;B}\, \Omega^{(m)}_{n,k}=0\,,\qquad \ell \in \mathbb{N}
 \;,
\end{equation}
for all level-$\ell$ generators of $Y\big(\mathfrak{gl}(m+k)\big)$. In particular, by using the Quantum Inverse Scattering Method, we will 
 relate the volume functions to the eigenvectors of the monodromy matrix of a particular spin chain, as in \cite{Kanning:2014maa}. Let us start by defining the latter. The quantum space of our spin chain is taken to be
\begin{equation}\label{quantum.space}
V=\overline{V}_1\otimes \ldots \otimes\overline{V}_k\otimes V_{k+1}\otimes\ldots\otimes V_n\,,
\end{equation}
where $\overline{V}_i$ and $V_i$ are representation spaces of particular $\mathfrak{gl}(m+k)$ representations. These representations are non-compact and elements of their representation spaces are functions, or more generally distributions, of bosonized momentum twistors. In the following we demand that the elements of the quantum space are invariant under rescaling of the variables $Z_i$. For $i=1,\ldots,k$, we identify pairs $\left(-\frac{\partial}{\partial Z^B_i},Z_i^A\right)$ as creation and annihilation operators acting on $\overline{V}_i$ and define the Fock vacua as
\begin{equation}
Z^A_i|\overline{0}\rangle_i=0\,,\quad A=1,\ldots m+k \qquad\Rightarrow \qquad |\overline{0}\rangle_i=\delta(Z_i^A)\,.
\end{equation}
These oscillator representations correspond to dual realizations in \cite{Kanning:2014maa}.
For $i=k+1,\ldots,n$, we identify pairs $\left(Z^A_i,\frac{\partial}{\partial Z^B_i}\right)$ as creation and annihilation operators acting on $V_i$ and define the Fock vacua as
\begin{equation}
\frac{\partial}{\partial Z^A_i}|0\rangle_i=0\,,\quad A=1,\ldots m+k \qquad\Rightarrow \qquad |0\rangle_i=1\,.
\end{equation}
These oscillator representations correspond to symmetric realizations in \cite{Kanning:2014maa}. We also equip each $V_i$ and $\overline{V}_i$ with a complex parameter $v_i$, called {\it inhomogeneity}, which will enter our construction as a parameter of Lax operators.
 The generators of the $\mathfrak{gl}(m+k)$ algebra are realized on these spaces as
\begin{equation}
(\overline{J}_i)^A_{\;B}=\frac{\partial}{\partial Z_i^B}Z_i^A =Z_i^A\frac{\partial}{\partial Z_i^B}+\delta_{B}^A\,\qquad i=1,\ldots k\,,
\end{equation}  
for all $\overline{V}_i$ and
\begin{equation}
(J_i)^A_{\;B}=Z_i^A \frac{\partial}{\partial Z_i^B}\,\qquad i=k+1,\ldots,n\,,
\end{equation}  
for all $V_i$. 

\begin{figure}[h!]
\centering
\def\svgwidth{7cm}
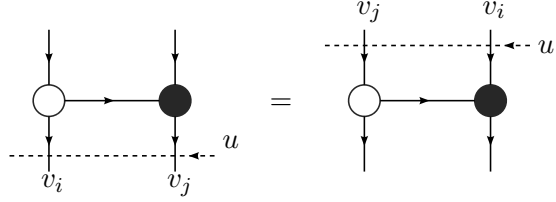
\caption{Fundamental relation for the construction of monodromy eigenvectors.}
\label{Fig.Intertwining}
\end{figure}

In addition to the quantum space $V$ of \eqref{quantum.space}, we introduce an auxiliary space $V_{\mathrm{aux}}$ which is a fundamental representation space of $\mathfrak{gl}(m+k)$. For every $i=1,\ldots,k$ we define the \emph{dual Lax operators} $\overline{L}_i : V_{\textrm{aux}} \otimes \overline{V}_i \to V_{\textrm{aux}} \otimes \overline{V}_i$, with matrix elements
\begin{equation}\label{dual.Lax.operators}
\overline{L}_i(u-v_i)^A_{\;B}=\delta^A_{\;B}+(u-v_i-1)^{-1} \frac{\partial}{\partial Z_i^B}\,Z_i^A\,,
\end{equation}
and for every $i=k+1,\ldots,n$ the \emph{symmetric Lax operators} $ L_i : V_{\textrm{aux}} \otimes  V_i \to V_{\textrm{aux}} \otimes  V_i$, with matrix elements
\begin{equation}\label{symmetric.Lax.operators}
L_i(u-v_i)^A_{\;B}=\delta^A_{\;B}+(u-v_i)^{-1} Z_i^A\frac{\partial}{\partial Z_i^B}\,.
\end{equation}
It will be useful in the following to introduce the operator
\begin{equation}
\mathcal{L}_i(u)^A_{\;B}=(u-v_i) \, \delta^A_{\;B}+Z_i^A\frac{\partial}{\partial Z_i^B}\,,
\end{equation}
which is related to our Lax operators by
\begin{equation}
\overline{L}_i(u)=(u-v_i-1)^{-1}\mathcal{L}_i(u),\qquad  L_i(u)=(u-v_i)^{-1}\mathcal{L}_i(u)\,.
\end{equation}
The monodromy matrix is defined as
\begin{equation}\label{monodromy}
M(u;v_1,\ldots, v_n)^A_{\;B} = \overline{L}_1(u-v_1)^A_{\;C_1}\ldots \overline{L}_k(u-v_k)^{C_{k-1}}_{\;C_k}L_{k+1}(u-v_{k+1})^{C_k}_{\;C_{k+1}}\ldots L_n(u-v_n)^{C_{n-1}}_{\;B}\,.
\end{equation}
In the following we will prove that (deformed) volume functions -- when acted upon with~$J_Y $ -- are (left) eigenvectors of the monodromy matrix \eqref{monodromy}. In order to do so, we use the expression for the volume function written in terms of $\mathcal{B}_{ij}(u)$ operators and act on it with $M(u;v_1,\ldots,v_n)$. Crucially, we will use the intertwining relation
\begin{equation}\label{intertwining.relations}
\mathcal{L}_i(u-v_i)\mathcal{L}_j(u-v_j)\mathcal{B}_{ij}(v_j-v_i)=\mathcal{B}_{ij}(v_j-v_i)\mathcal{L}_i(u-v_j)\mathcal{L}_j(u-v_i)\,,
\end{equation}
which has been proven in \cite{Kanning:2014maa} and for which we provide details in Appendix \ref{app.intertwining}. Its meaning is depicted in Figure \ref{Fig.Intertwining}, where the dashed line corresponds to the auxiliary space and $\mathcal{B}_{ij}$ is again represented by a composition of one black and one white trivalent vertex. 

We proceed in full generality, keeping the deformation parameters $u$ in \eqref{Boperator} non-zero. We use the permutation we introduced in \eqref{topcellsigma} with decomposition \eqref{topcelldecomposition} and  define a sequence of permutations
\begin{equation}
\tau_l=\tau_{l-1}\circ (i_l,j_l)=(i_1 j_1)\ldots (i_l,j_l)\,,
\end{equation}
with the property
\begin{equation}
\tau_{k(n-k)}=\sigma_{n,k}^{-1}\,.
\end{equation}
Then we can define a deformed volume function as
\begin{equation}\label{omega.from.seed.genericu}
\Omega_{n,k}^{(m)}(Y,Z;v_1,\ldots,v_n)=\mathcal{B}_{i_1j_1}(\bar u_1)\,\mathcal{B}_{i_2j_2}(\bar u_2)\ldots \mathcal{B}_{i_{k(n-k)}j_{k(n-k)}}(\bar u_{k(n-k)})\,\mathcal{S}_k^{(m)} \,,
\end{equation}
with 
\begin{equation}
\bar u_l=v_{\tau_l(i_l)}-v_{\tau_l(j_l)}\,.
\end{equation}
In order for \eqref{omega.from.seed.genericu} to be an element of our quantum space $V$, $\Omega_{n,k}^{(m)}$ must be invariant under rescalings of the $Z_i$, which requires the following restriction on the inhomogeneities $v_i$:
\begin{equation}\label{vrelation}
v_{\sigma_{n,k}(i)}=v_i\,,
\end{equation}
\textit{i.e.} a particular case of the condition derived in \cite{Kanning:2014maa}, see also \cite{Broedel:2014pia}. Then we can prove the following relation
\begin{equation}\label{monodromy.invariance}
 M(u;v_1,\dots,v_n)^C_B\, (J_Y)_C^A \,\Omega_{n,k}^{(m)} = (J_Y)_B^A \,\Omega_{n,k}^{(m)} \,,
\end{equation}
namely, $J_Y \,\Omega_{n,k}^{(m)}$ is an eigenvector of the monodromy matrix \eqref{monodromy} with unit eigenvalue. In particular, if we take the undeformed case, that is $v_i=0$ for all $i$, then \eqref{vrelation} is trivially satisfied and \eqref{monodromy.invariance} holds true. This will lead us in the following section to the meaning of Yangian invariance for the amplituhedron volume functions.

\subsection{Proof of the Monodromy Relation}

In the following we provide the main steps to prove \eqref{monodromy.invariance}, which is the most important formula of the paper. In order to ease the notation, we introduce an auxiliary object:
\begin{equation}
   \mathcal{M}(u;v_1,\ldots,v_n) := {\mathcal L}_1(u-v_1) \ldots {\mathcal L}_n(u-v_n) \, .
\end{equation}
Then the monodromy matrix can be rewritten as 
\begin{equation}
M(u;v_1,\ldots,v_n) =
 \prod_{i=1}^k \frac1{u-v_i-1}\prod_{i=k+1}^n \frac1{u-v_i} \mathcal{M}(u;v_1,\ldots,v_n) \, .
\end{equation}
We will need a technical result, whose proof we postpone to Appendix \ref{technical.lemma.proof}:
\begin{equation}\label{technical.lemma}
 {\mathcal L}_i(u,v_i)^C_B\, (J_Y)^A_C  \,\mathcal S_k^{(m)} = 
\begin{cases} (u-v_i-1)\, (J_Y)^A_B \,\mathcal S_k^{(m)} \, ,\quad i=1,\dots,k \,\\
(u-v_i)\, (J_Y)^A_B \,\mathcal S_k^{(m)} \, ,\quad\quad\;\;\, i=k+1,\dots,n \,.
\end{cases}
\end{equation}
Using this fact, we can show how the monodromy matrix acts on the seed function $\mathcal{S}_k^{(m)}$:
\begin{align}\label{monodromy.on.seed}
 \mathcal M(u;v_1,\ldots,v_n)^C_{\;B} \,(J_Y )^A_{\;C}\, \mathcal{S}_k^{(m)} &= 
 ({\mathcal L}_1)^C_{\;D_1} \ldots ({\mathcal L}_n)^{D_{n-1}}_{\;B}\,(J_Y)^A_{\;C}  \,\mathcal S_k^{(m)} \nonumber \\
&= \prod_{i=1}^k (u-v_i-1) \prod_{i=k+1}^n (u-v_i) \,(J_Y)^A_{\;B} \, \mathcal S_k^{(m)} \, ,
\end{align}
where we have suppressed all arguments of the Lax operators and we have made repeatedly use of \eqref{technical.lemma}. We are now ready to prove that $J_Y \,\Omega_{n,k}^{(m)}$ is indeed an eigenvector of our monodromy matrix. Following the same steps as in \cite{Kanning:2014maa} and using the intertwining relations \eqref{intertwining.relations}, we have
\begin{align}
&M(u;v_1,\dots,v_n)_{\;B}^C\,(J_Y)^A_{\;C}   \,\Omega_{n,k}^{(m)} = \\
&=\prod_{i=1}^k \frac1{u-v_i-1}\prod_{i=k+1}^n \frac1{u-v_i}  \mathcal{M}(u;v_1,\dots,v_n)^C_{\;B}\,\prod_{l=1}^{k(n-k)} \mathcal B_{i_l,j_l}(\bar u_l) \,(J_Y)_{\;C}^A\, \mathcal S_k^{(m)}  \nonumber \\
&= \prod_{i=1}^k \frac1{u-v_i-1} \prod_{i=k+1}^n \frac1{u-v_i} \prod_{l=1}^{k(n-k)} \mathcal B_{i_l,j_l}(\bar u_l)   \mathcal{M}(u;v_{\sigma_{n,k}^{-1}(1)},\dots,v_{\sigma_{n,k}^{-1}(n)})^C_{\;B} \,(J_Y)_{\;C}^A \,\mathcal S_k^{(m)} \,,
\end{align}
Observe that $J_Y$ commutes with all the $\mathcal B_{i_l j_l}(\bar u_l)$, since the latter depend on $Z_i$ alone. Now, using formula \eqref{monodromy.on.seed}, we finally arrive at the desired result
\begin{align}
&M(u;v_1,\dots,v_n)_{\;B}^C \,(J_Y)^A_{\;C}  \,\Omega_{n,k}^{(m)}= \\
&=\prod_{i=1}^k \frac{u-v_{\sigma_{n,k}^{-1}(i)}-1}{u-v_i-1} \prod_{i=k+1}^n \frac{u-v_{\sigma_{n,k}^{-1}(i)}}{u-v_i} \prod_{l=1}^{k(n-k)} \mathcal B_{i_l,j_l}(\bar u_l) (J_Y)^A_{\;B} \,\mathcal S_k^{(m)} = (J_Y)^A_{\;B} \,\Omega_{n,k}^{(m)} \,,
\end{align}
where the products over $i$ evaluate to 1 in light of \eqref{vrelation}. This completes the proof of formula \eqref{monodromy.invariance}.


\section{Yangian Invariance for the Amplituhedron Volume}
\label{invariance}

We are now ready to discuss the Yangian invariance of the amplituhedron and, specifically, of the volume function.
By defining 
\begin{equation}\label{omega.A.B}
\Omega^A_{\;B}(Y,Z;v_1,\dots,v_n) := (J_Y)^A_{\;B} \, \Omega_{n,k}^{(m)}(Y,Z;v_1,\ldots,v_n) \,,
\end{equation}
we can rewrite the result \eqref{monodromy.invariance} of the previous section in the following way:
\begin{equation}\label{monodromy.omega}
M(u;v_1,\ldots,v_n)^C_{\;B} \,\Omega^A_{\;C}(Y,Z;v_1,\dots,v_n) = \Omega^A_{\;B}(Y,Z;v_1,\dots,v_n) \, .
\end{equation}
If we now expand the monodromy matrix around $u\to \infty$ using the explicit form of Lax operators \eqref{dual.Lax.operators} and \eqref{symmetric.Lax.operators}, we find
\begin{equation}
M(u;v_1,\ldots,v_n)^A_{\;B}=\delta^A_{\;B} + \frac{1}{u}\,(J^{(0)})^A_{\;B} + \,\frac{1}{u^2}(J^{(1)})^A_{\;B} + \ldots\,.
\end{equation}
In particular, the leading term cancels the right hand side of \eqref{monodromy.omega} while the subleading terms lead to the following invariance properties for $\Omega^A_{\;B}$:
\begin{equation}\label{level.zero.one.invariance}
(J^{(0)})^C_{\;B}\, \Omega^A_{\;C} = 0 \, ,\qquad (J^{(1)})^C_{\;B} \,\Omega^A_{\;C} = 0 \,.
\end{equation}
Setting the inhomogeneities $v_i$ to zero for compactness, we find
\begin{align}\label{level0.generators}
(J^{(0)})^A_{\;B} &=\sum_{i=1}^n Z_i^A\frac{\partial}{\partial Z^B_i}+k\,  \delta^A_{\;B}\,,\\ \label{level1.generators}
(J^{(1)})^A_{\;B} &=\sum_{i<j} Z_i^A\frac{\partial}{\partial Z^C_i}Z_j^C\frac{\partial}{\partial Z^B_j} + k \sum_{i=1}^n  Z_i^A\frac{\partial}{\partial Z^B_i}+\frac{k(k+1)}{2}\delta^A_{\;B} \,.
\end{align}
In order to make a comparison with formulas already present in the literature, one can use the level-zero invariance and rewrite the level-one generators in the form
\begin{equation}
(J^{(1)})^A_{\;B}=\sum_{i<j} \left({Z}_i^{A}\frac{\partial}{\partial {Z}^{C}_i}{Z}_j^{C}\frac{\partial}{\partial {Z}^{B}_j}-(i\leftrightarrow j)\right) \,.
\end{equation}
These generators are known to form the Yangian algebra $Y\big(\mathfrak{gl}(m+k)\big)$. We have therefore shown that the functions $\Omega^A_{\;B}$, related to our original volume function  $\Omega_{n,k}^{(m)}$ through \eqref{omega.A.B},  are Yangian-invariant. 

As a final remark, let us mention that at the moment it is not clear what is the explicit relation, if any, between the Yangian for the amplituhedron with $m=4$, namely $Y\big(\mathfrak{gl}(4+k)\big)$, and the Yangian for scattering amplitudes, namely $Y\big(\mathfrak{psl}(4|4)\big)$. In particular, it is not known how to directly translate the bosonic generators \eqref{level0.generators} and \eqref{level1.generators} by integrating out the auxiliary fermions $\phi$, as in \eqref{from.volume.to.amplitude}, in order to get \eqref{yangiangenampls}. It is not even clear whether the bosonic part of $Y\big(\mathfrak{psl}(4|4)\big)$ can be embedded in a larger $Y\big(\mathfrak{gl}(4+k)\big)$. The statement we proved in this paper is that all volume functions corresponding to Yangian invariants of $Y\big(\mathfrak{psl}(4|4)\big)$ are $Y\big(\mathfrak{gl}(4+k)\big)$ Yangian invariant as in \eqref{level.zero.one.invariance}. However, we can easily understand  that the $\mathfrak{u}(1)$ part of $\mathfrak{gl}(4+k)$ corresponds to the central charge $\mathfrak{C}$ of $\mathfrak{psl}(4|4)$ since both provide the scaling properties of the final result.

\subsection{Relation to Positive Diffeomorphisms}
As pointed out by the authors of \cite{ArkaniHamed:2012nw,*ArkaniHamed:2012nwB}, the Yangian invariance of scattering amplitudes is strictly related to the diffeomorphisms of the positive Grassmannian -- or \emph{positive} diffeomorphisms. It was indeed shown that their infinitesimal generators match the level-one Yangian generators of \cite{Drummond:2010qh}. 
Following their steps, we want to show that our Yangian generators are related to diffeomorphisms as well. There is, however, a crucial difference compared to the scattering amplitudes case: we need to be precise when we talk about $\Omega_{n,k}^{(m)}(Y,Z)$ as a Grassmannian integral, since the integrand of \eqref{omegaintegral} is not invariant under the  $GL(k)$-transformations performed on the matrix $C=(c_{\alpha i})$. Instead, in order to have a well-defined integral on the Grassmannian, we need to work with the volume form \eqref{volume.form}.
Then, when we consider diffeomorphisms of the Grassmannian space $G_+(k,n)$, we need to always supplement them by a \mbox{$GL(k)$-transformation} acting on the $Y$-space. A generic infinitesimal diffeomorphism for $c_i$ and a $GL(k)$-transformation for $Y$ take the form
\begin{align}
c_{ i} &\rightarrow c_{ i} +\delta c_i \,,\\
Y^A_\alpha &\rightarrow Y^A_\alpha + y^{\;\beta}_\alpha \, Y^A_\beta+\ldots\,,
\end{align}
where $y^{\;\beta}_\alpha$ parametrizes an infinitesimal $\mathfrak{gl}(k)$ transformation. We focus here on positive diffeomorphisms which implies that $\delta c_i$ should preserve the positive stratification of the Grassmannian. In that case we can use a slightly modified version of the argument in \cite{ArkaniHamed:2012nw,*ArkaniHamed:2012nwB}: in order to preserve the $GL(k)$-invariance of \eqref{volume.form} the leading diffeomorphisms have to be of the following form
\begin{equation}
\delta c_{\alpha i} = y^{\;\beta}_\alpha c_{\beta i} \,,
\end{equation}
with the same matrix $y$.
We also find the second order diffeomorphisms which combine into
\begin{equation}
\delta c_i  = y \cdot c_i + \sum_{j<i} (c_j + y\cdot c_j) (\omega_j \cdot c_i)\,,
\end{equation}
with constraint
\begin{equation}
\sum_i \left(c_{\alpha i} +  y^{\;\beta}_\alpha  c_{\beta i}\right) \omega^{\;\gamma}_i = 0 \,, \quad \forall \, \alpha, \gamma   \,.
\end{equation}
Using these diffeomorphisms and expanding \eqref{volume.form}, we find that the term proportional to $y$ and not quadratic in the $c_i$ cancels and we are left with
\begin{equation}\label{diffeos.second}
\sum_{j<i} \left(c_{\alpha j} +  y^{\;\beta}_\alpha  c_{\beta j} \right) \left(\omega^{\;\gamma}_j  c_{\gamma i}\right) \frac{\partial}{\partial c_{\alpha i}} \,.
\end{equation}
This agrees with the diffeomorphisms found in \cite{ArkaniHamed:2012nw,*ArkaniHamed:2012nwB} if $y^{\; \beta}_\alpha = 0$. It was shown there that \eqref{diffeos.second} with $y^{\;\beta}_\alpha=0$ can be related to level-one Yangian generators of the form
\begin{equation}
\sum_{i<j} \left( Z_i^A\frac{\partial}{\partial Z^C_i}Z_j^C\frac{\partial}{\partial Z^B_j} \right)\,.
\end{equation} 
On the other hand, the $y$-dependent part of \eqref{diffeos.second} can be shown to be related to the following third-order differential operator
\begin{equation}
\sum_{i<j} \sum_\alpha \left(Z_i^D\frac{\partial}{\partial Z^C_i}Z_j^C\frac{\partial}{\partial Z^B_j}Y_\alpha^A\frac{\partial}{\partial Y^D_\alpha} \right) \,.
\end{equation}
We have previously established that a particular combination of these two operators annihilates the volume function, see \eqref{level.zero.one.invariance} together with \eqref{omega.A.B} and \eqref{operator.JY}. However, we could not fix such combination directly from the positive diffeomorphisms analysis.


\section{Conclusions and Outlook}

In this paper we have explored the Yangian symmetry of the amplituhedron at tree-level. We have shown that it admits a natural description in terms of integrable spin chains. In particular, we proved that the volume function can be constructed as in \eqref{omega.from.seed} and that a suitable modification of it \eqref{omega.A.B} is invariant under the Yangian of the $\mathfrak{gl}(m+k)$ algebra. This descends from the monodromy condition \eqref{monodromy.invariance}. Furthermore, we have introduced a new on-shell diagrammatics, based on three types of vertices: two of them are a slight modification of the standard trivalent ones, whereas the new seed vertex \eqref{seed.vertex} accounts for the auxiliary $k$-plane $Y$. We showed how to classify such diagrams by extending the cluster structure of on-shell diagrams for amplitudes to include also the seed vertex.

Based on our results, one could try to answer several questions. The most pressing one concerns our previous work \cite{Ferro:2015grk}, where the Capelli differential equations -- supplemented by covariance and scaling conditions -- were in general only partially constraining the ansatz we proposed for the volume functions, written as an integral over a dual space. As suggested there, the additional conditions coming from Yangian invariance should completely fix, or at least further constrain such ansatz. We leave it for future work. 
Moreover, the integrability approach might help us to get a handle on loop-level volume forms as well. It has been observed in \cite{ArkaniHamed:2010kv} that all integrands of scattering amplitudes are Yangian-invariant, nevertheless the analogous statement for the  loop-level volume functions was beyond reach until now.
In \cite{Bai:2015qoa} it was shown that there exists a generalization of the Grassmannian measure and of the cell structure for one-loop volume functions. It would be interesting to explore possible implications of our present results in these respects.


\section*{Acknowledgements}
We would like to thank Nima Arkani-Hamed, Jacob Bourjaily and Nils Kanning for very useful discussions. L.F. is supported by the Elitenetwork of Bavaria. T.L. is supported by ERC STG grant 306260. The work of M.P. is funded by QMUL Principal's Studentship. This work was partially supported by the DFG Grant FE 1529/1-1.


\appendix
\addtocontents{toc}{\protect\setcounter{tocdepth}{0}}

\section{Construction of Volume Functions}\label{app.omega.proof}

We will prove the identity \eqref{omega.from.seed}
\begin{equation}\label{omega.from.seed.app}
\Omega_{n,k}^{(m)}(Y,Z) = \prod_{l=1}^{k(n-k)}\mathcal{B}_{i_l j_l}(0)\, \mathcal{S}_k^{(m)} \, .
\end{equation}
First of all, we know \cite{Kanning:2014maa} that in the case of amplitudes 
\begin{equation}\label{B.on.deltas}
\prod_{l=1}^{k(n-k)}\mathcal{B}_{i_l j_l}(0) \prod_{i=1}^k \delta^{4|4}(\mathcal{Z}^\mathcal{A}_i) = \bigintsss \frac{d^{k\cdot(n-k)}\tilde c}{\prod_{a=1}^n \widetilde{\mathcal M}_a} \; \delta^{4|4}(\tilde C \cdot \mathcal Z) \, ,
\end{equation}
{\it i.e.}~after changing variables, we can recover a gauge-fixed version of the Grassmannian formula \eqref{grass.integral}, with $\tilde C = \big( \mathbb I_{k \times k} \big | F_{k\times(n-k)}\big)$ being a matrix with the first $k$ columns fixed to the identity and $\widetilde{\mathcal M}_a$ its consecutive maximal minors. We can now use this result and act with the $\mathcal B_{i_l j_l}(0)$ operators on the seed $\mathcal S_k^{(m)}$ instead of the $\delta$-functions:
\begin{equation}\label{gaugefixed.Grassmannian.integral}
\begin{split}
\prod_{l=1}^{k(n-k)}\mathcal{B}_{i_l j_l}(0) \,\mathcal S_k^{(m)} &= \bigintsss \frac{d^{k \cdot k}\beta}{(\det \beta)^k} \prod_{l=1}^{k(n-k)}\mathcal{B}_{i_l j_l}(0) \, \delta^{k(m+k)}(Y - \beta \cdot Z) = \\
&= \bigintsss \frac{d^{k \cdot k}\beta}{(\det \beta)^k} \bigintsss \frac{d^{k\cdot(n-k)} \tilde c}{\prod_{a=1}^n \widetilde{\mathcal M}_a} \; \delta^{k(m+k)}(Y - \beta \cdot \tilde C \cdot Z) \, .
\end{split}
\end{equation}
The amplituhedron volume function can be reconstructed by the following change of variables: $\beta' = \beta$ and $F' = \beta \cdot F$, such that the new variables are rearranged in the matrix $C = \big( \beta' \big| F' \big)=\beta \cdot \tilde C $. The related Jacobian is $(\det \beta)^{-(n-k)}$, whereas each of the $n$ minors $\widetilde{\mathcal M}_a$ of $\tilde C$ equals the corresponding $\mathcal M_a$ of $C$, up to a factor $(\det \beta)^{-1}$. In the end,
\begin{equation}
\prod_{l=1}^{k(n-k)}\mathcal{B}_{i_l j_l}(0) \,\mathcal S_k^{(m)} = \bigintsss \frac{d^{k \cdot n} c}{\prod_{a=1}^n \mathcal M_a} \; \delta^{k(m+k)}(Y - C \cdot Z) = \Omega_{n,k}^{(m)} \, .
\end{equation}

\section{Intertwining Relations}\label{app.intertwining}
In this appendix we will prove formula \eqref{intertwining.relations}, assuming $i \not = j$
\begin{equation}\label{intertwining.relations.App}
\mathcal{L}_i(u-v_i)\mathcal{L}_j(u-v_j)\mathcal{B}_{ij}(v_j-v_i)=\mathcal{B}_{ij}(v_j-v_i)\mathcal{L}_i(u-v_j)\mathcal{L}_j(u-v_i)\,.
\end{equation}
We observe that on both sides we get the same contribution proportional to $\delta^A_{\;B}$, namely 
\begin{equation}
(u-v_i)(u-v_j)\mathcal{B}_{ij}(v_j-v_i) \, .
\end{equation}
Focusing on the other terms, we have (denoting $\nu_{i} := u-v_i$ and $\nu := \nu_i-\nu_j=v_j-v_i$)
\begin{align} \label{IntRel.LHS}
\left( \nu_{i} \, Z_j^A\frac{\partial}{\partial Z_j^B}+ \right. & \left.\nu_{j} \, Z_i^A\frac{\partial}{\partial Z_i^B}+ Z_i^A\frac{\partial}{\partial Z_i^C} Z_j^C\frac{\partial}{\partial Z_j^B} \right) \left( Z_j^D\frac{\partial}{\partial Z_i^D}\right)^{\nu} = \nonumber \\
&= \left( Z_j^D\frac{\partial}{\partial Z_i^D}\right)^{\nu} \left( \nu_{j} \, Z_j^A\frac{\partial}{\partial Z_j^B}+\nu_{i} \, Z_i^A\frac{\partial}{\partial Z_i^B}+ Z_i^A\frac{\partial}{\partial Z_i^C} Z_j^C\frac{\partial}{\partial Z_j^B} \right) 
\,.
\end{align}
First, we observe that
\begin{equation}
\left[ Z_l^A\frac{\partial}{\partial Z_l^B},Z_j^D\frac{\partial}{\partial Z_i^D } \right]=\left( \delta_{jl}-\delta_{il} \right) Z_j^A\frac{\partial}{\partial Z_i^B} \,.
\end{equation}
The RHS trivially commutes with $Z_j^D\frac{\partial}{\partial Z_i^D } $, thus we can use the formula
\begin{equation}
\big[[A,B],B \big]=0 \Rightarrow [A,B^\nu]=\nu[A,B]B^{\nu-1} \,.
\end{equation}
to perform the commutation in the first two summands of the LHS of \eqref{IntRel.LHS}, obtaining
\begin{equation}
\nu(\nu_i-\nu_j)Z_j^A\frac{\partial}{\partial Z_i^B} \left( Z_j^D\frac{\partial}{\partial Z_i^D}\right)^{\nu-1} \, ,
\end{equation} 
and in the third one, after some manipulations, arriving at
\begin{equation}
\nu\left( Z_j^D\frac{\partial}{\partial Z_i^D}\right)^{\nu} \left( Z_i^A\frac{\partial}{\partial Z_i^B}-Z_j^A\frac{\partial}{\partial Z_j^B}\right)-\nu^2\left( Z_j^D\frac{\partial}{\partial Z_i^D}\right)^{\nu-1} Z_j^A\frac{\partial}{\partial Z_i^B} \,.
\end{equation}
Now it is immediate to check that \eqref{IntRel.LHS} turns into
\begin{equation}
(\nu+\nu_j-\nu_i) \left[ \left( Z_j^D\frac{\partial}{\partial Z_i^D}\right)^{\nu} \left( Z_i^A\frac{\partial}{\partial Z_i^B}-Z_j^A\frac{\partial}{\partial Z_j^B}\right) - \nu \left( Z_j^D\frac{\partial}{\partial Z_i^D}\right)^{\nu-1} Z_j^A\frac{\partial}{\partial Z_i^B} \right] = 0 \,,
\end{equation}
which holds true since $\nu = \nu_i - \nu_j$.

\section{Action of the Monodromy Matrix on the Seed}\label{technical.lemma.proof}

We will prove the identity \eqref{technical.lemma}
\begin{equation}\label{technical.lemma.oncemore}
 {\mathcal L}_i(u,v_i)^C_B\,  \,(J_Y)^A_C \,\mathcal S_k^{(m)} = 
\begin{cases} (u-v_i-1)\, (J_Y)^A_B \,\mathcal S_k^{(m)} \, ,\quad i=1,\dots,k \,\\
(u-v_i)\, (J_Y)^A_B \,\mathcal S_k^{(m)} \, ,\quad i=k+1,\dots,n \,.
\end{cases}
\end{equation}
In the following we will work with the matrix elements and suppress the arguments of the ${\mathcal L}_i$ operators.  We start with
\begin{align}
 {\mathcal L}_i(u,v_i)^C_B\, (J_Y)^A_C\mathcal S_k^{(m)}&=
 \bigg((u-v_i)\delta^C_{\;B} + Z_i^C\frac{\partial}{\partial Z_i^B}\bigg) \bigg(\sum_{\alpha=1}^k Y_\alpha^A\frac{\partial}{\partial Y_\alpha^C} +k \delta^A_{\;C}\bigg) \mathcal S_k^{(m)} = \nonumber \\
&= (u-v_i) (J_Y)^A_{\;B} \,\mathcal S_k^{(m)} + Z_i^C \frac{\partial}{\partial Z_i^B} \bigg(\sum_{\alpha=1}^k Y_\alpha^A\frac{\partial}{\partial Y_\alpha^C} +k \delta^A_{\;C}\bigg) \mathcal S_k^{(m)} \,.
\end{align}
We argue that the second summand is trivially zero for $i>k$, since $\mathcal S_k^{(m)}$ just depends on $Z_1,\dots,Z_k$, whereas it is equal to $-(J_Y)^A_B \,\mathcal S_k^{(m)} $ for $i=1,...,k$. Using the fact that the seed $\mathcal S_k^{(m)}$ is $GL(m+k)$-covariant in the same way as $\Omega^{(m)}_{n,k}$ is, see \cite{Ferro:2015grk}, we can write the second summand as
\begin{equation}
- Z_i^C \frac{\partial}{\partial Z_i^B} \bigg(\sum_{l=1}^k Z_l^A\frac{\partial}{\partial Z_l^C}  \bigg) \mathcal S_k^{(m)} \,.
\end{equation}
Now we recast the operators acting on $Z_i$ into ones acting on the integration variables $\beta_{\alpha i}$:
\begin{align} \label{AppCref1}
\sum_{l=1}^k Z_i^C \frac{\partial}{\partial Z_i^B} Z_l^A \frac{\partial}{\partial Z_l^C} &= \sum_{l=1}^k \frac{\partial}{\partial Z_i^B} Z_l^A Z_i^C \frac{\partial}{\partial Z_l^C} -\sum_{l=1}^k Z_l^A \frac{\partial}{\partial Z_l^B} \,,
\end{align}
finally noticing that 
\begin{equation}
Z_i^C \frac{\partial}{\partial Z_l^C} \,\mathcal S_k^{(m)} = \int \frac{d^{k \cdot k}\beta}{(\det \beta)^k}\, \mathbb O^i_l \,\delta^{k(m+k)}(Y-\beta \cdot Z) \,,\quad \mathbb O^i_l := \beta_{\alpha l}\frac{\partial}{\partial \beta_{\alpha i}} \,.
\end{equation}
Suppressing the argument of the $\delta$-function,
\begin{align}
\int \frac{d^{k \cdot k}\beta}{(\det \beta)^k} &\, \mathbb O^i_l \,\delta^{k(m+k)} = \int d^{k \cdot k}\beta \bigg( \mathbb O^i_l (\det \beta)^{-k} - [\mathbb O^i_l, (\det \beta)^{-k}]\bigg) \delta^{k(m+k)} = \nonumber \\
& \hspace{-1.5cm}= \int d^{k \cdot k}\beta \bigg( \frac{\partial}{\partial \beta_{\alpha i}} \beta_{\alpha l} - k\, \delta^i_l \bigg) (\det \beta)^{-k} \, \delta^{k(m+k)} +  k \, \delta^i_l \int d^{k \cdot k}\beta (\det \beta)^{-k} \,\delta^{k(m+k)} = 0 \,,
\end{align}
where in the first integral we have dropped the total derivative term and in the second one we used
\begin{equation*}
[\mathbb O^i_l, (\det \beta)^{-k}] = -k \, (\det \beta)^{-k-1} [\mathbb O^i_l, \det \beta] = -k \, \delta^i_l (\det \beta)^{-k} \,,
\end{equation*}
since the operator $\mathbb O^i_l$ acts on the determinant simply substituting the row $i$ with the row $l$.
Substituting back all the intermediate results and using again $GL(m+k)$ covariance of the seed to rewrite the second term in the LHS of \eqref{AppCref1}, we readily obtain the desired identity \eqref{technical.lemma.oncemore}.

\section{Seed Cluster Mutation}\label{App.seed.mutation}

\begin{figure}[htb] 
  \centering
  \def\svgwidth{300pt}
  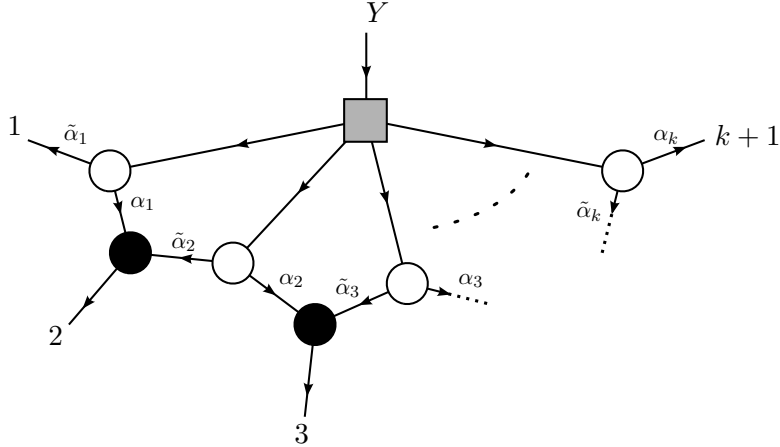
  \caption{On-shell diagram}
  \label{Fig.perfectorientation}
\end{figure}

In this appendix we prove that the seed cluster mutations depicted in Fig.~\ref{Fig.ClusterMutations} hold true for any $k$. We consider the on-shell diagram portrayed in Figure \ref{Fig.perfectorientation} and show that we can cyclically relabel particles $1,\ldots,k+1$ and the diagram evaluates to the same Grassmannian integral. Using the rules of Figure \ref{Fig.(anti)MHV3} and solving the $\delta$-functions arising from the trivalent vertices, we obtain an integral representation involving both $\alpha, \tilde\alpha$ and $\gamma,\tilde\gamma$ variables, respectively associated to white and black vertices. The latter can be eliminated fixing $2(k-1)$ of the $GL(1)$ redundancies due to the projective nature of the bosonized twistors running along the internal lines and we arrive at
\begin{equation} \label{alphatildebeta}
\bigintsss \prod_{a=1}^{k} \frac{d \alpha_a}{\alpha_a} \prod_{a=1}^{k} \frac{d \tilde{\alpha}_a}{\tilde{\alpha}_a}   \bigintsss \frac{d^{k\cdot k}\beta}{(\det \beta)^k} \; \delta^{k(m+k)}(Y - \beta \cdot  C \cdot Z) \; ,
\end{equation}
where $C$ is a matrix whose entries read $C(\tilde\alpha,\alpha)_{ai}=\delta_{ai} \tilde{\alpha}_a + \delta_{a, i-1} \alpha_{a}$. Using the last $k$ $GL(1)$ redundancies, we can eliminate the $\tilde\alpha$ variables too, which yields an even simpler form for $C$, namely $C(\alpha)_{ai}=\delta_{ai} + \delta_{a, i-1} \alpha_{k+1-a}$. In formulas, 
\begin{equation}
C(\alpha,\tilde\alpha) = \begin{pmatrix}
\tilde\alpha_1 	& \alpha_1	& 			& 			&  	\\
			& \tilde\alpha_2 & \alpha_2	& 			&  	\\
			&			& \ddots		&			& 	\\
			& 			&			& \tilde\alpha_k	& \alpha_k
\end{pmatrix} \quad\longrightarrow\quad
C(\alpha) = \begin{pmatrix}
1 	& \alpha_1& 			& 		&  	\\
	& 1 		& \alpha_2	& 		&  	\\
	&		& \ddots		&		& 	\\
	&		&			& 1		& \alpha_k
\end{pmatrix} \;,
\end{equation}
resulting in the following expression for the on-shell diagram of Figure \ref{Fig.perfectorientation}:
\begin{equation} \label{alphabeta}
\bigintsss \prod_{a=1}^k \frac{d \alpha_a}{\alpha_a} \bigintsss \frac{d^{k\cdot k}\beta}{(\det \beta)^k} \; \delta^{k(m+k)}(Y - \beta \cdot  C(\alpha) \cdot Z) \;.
\end{equation}

We would like to give $C$ an even simpler structure, such that its first $k$ columns form an identity matrix and only the last one is non-trivial. This is achieved using the following $GL(k)$ transformation: if $A$ is the square matrix formed by the first $k$ columns of $C$, then $A^{-1}C=(\mathbb{I}_{k \times k}| F_{k \times 1})$, with $F(\alpha)_{a}=(-1)^{k-a} \alpha_1...\alpha_{k+1-a}$.
At this point we change variables from $\alpha$'s to $c$'s, namely $c_{a, k+1}=F_a(\alpha)$, and from $\beta_{aj}$ to $\beta'_{ab}=(\beta \cdot A)_{ab}$. Calling $\widetilde{\mathcal M}_i$ the minor $(i \ldots i+k-1)$ of the matrix $\tilde{C}=(\mathbb{I}_{k \times k}|c_{a,k+1})$, we observe that
\begin{equation}
\prod_{a=1}^k \frac{d \alpha_a}{\alpha_a}=\prod_{a=1}^k \frac{d c_{a, k+1}}{c_{a, k+1}}=\frac{d^k c}{\prod_{i=1}^{k+1} \widetilde{\mathcal M}_i} \;,
\end{equation}
and we recover the gauge-fixed version of the Grassmannian integral over $C$ already appeared in \eqref{gaugefixed.Grassmannian.integral}. Following the same steps presented in Appendix A, we can thus bring \eqref{alphabeta} to the form
\begin{equation}
\bigintsss \frac{d^{k(k+1)} c}{\prod_{a=1}^{k+1} \mathcal M_a} \; \delta^{k(m+k)}(Y - C \cdot Z) = \Omega_{k+1,k}^{(m)}(Y,Z) \;,
\end{equation} 
which exhibits a manifest cyclic symmetry in $Z_1, \dots, Z_{k+1}$.

\bibliographystyle{nb}
\bibliography{Yangian}

\end{document}

%% file: Images/White3vertex_new.pdf_tex
\begingroup%
  \makeatletter%
  \providecommand\color[2][]{%
    \errmessage{(Inkscape) Color is used for the text in Inkscape, but the package 'color.sty' is not loaded}%
    \renewcommand\color[2][]{}%
  }%
  \providecommand\transparent[1]{%
    \errmessage{(Inkscape) Transparency is used (non-zero) for the text in Inkscape, but the package 'transparent.sty' is not loaded}%
    \renewcommand\transparent[1]{}%
  }%
  \providecommand\rotatebox[2]{#2}%
  \ifx\svgwidth\undefined%
    \setlength{\unitlength}{148.22399902bp}%
    \ifx\svgscale\undefined%
      \relax%
    \else%
      \setlength{\unitlength}{\unitlength * \real{\svgscale}}%
    \fi%
  \else%
    \setlength{\unitlength}{\svgwidth}%
  \fi%
  \global\let\svgwidth\undefined%
  \global\let\svgscale\undefined%
  \makeatother%
  \begin{picture}(1,1.28948628)%
    \put(0,0){\includegraphics[width=\unitlength]{White3vertex_new.pdf}}%
    \put(-0.04055306,0.66669578){\color[rgb]{0,0,0}\makebox(0,0)[lb]{\smash{$1$}}}%
    \put(1.00651086,1.12006366){\color[rgb]{0,0,0}\makebox(0,0)[lb]{\smash{$3$}}}%
    \put(1.00651086,0.10538317){\color[rgb]{0,0,0}\makebox(0,0)[lb]{\smash{$2$}}}%
  \end{picture}%
\endgroup%

%% file: Images/Black3vertex_new.pdf_tex
\begingroup%
  \makeatletter%
  \providecommand\color[2][]{%
    \errmessage{(Inkscape) Color is used for the text in Inkscape, but the package 'color.sty' is not loaded}%
    \renewcommand\color[2][]{}%
  }%
  \providecommand\transparent[1]{%
    \errmessage{(Inkscape) Transparency is used (non-zero) for the text in Inkscape, but the package 'transparent.sty' is not loaded}%
    \renewcommand\transparent[1]{}%
  }%
  \providecommand\rotatebox[2]{#2}%
  \ifx\svgwidth\undefined%
    \setlength{\unitlength}{148.22399902bp}%
    \ifx\svgscale\undefined%
      \relax%
    \else%
      \setlength{\unitlength}{\unitlength * \real{\svgscale}}%
    \fi%
  \else%
    \setlength{\unitlength}{\svgwidth}%
  \fi%
  \global\let\svgwidth\undefined%
  \global\let\svgscale\undefined%
  \makeatother%
  \begin{picture}(1,1.28948628)%
    \put(0,0){\includegraphics[width=\unitlength]{Black3vertex_new.pdf}}%
    \put(-0.04055306,0.66669578){\color[rgb]{0,0,0}\makebox(0,0)[lb]{\smash{$1$}}}%
    \put(1.00651086,1.12006366){\color[rgb]{0,0,0}\makebox(0,0)[lb]{\smash{$3$}}}%
    \put(1.00651086,0.10538317){\color[rgb]{0,0,0}\makebox(0,0)[lb]{\smash{$2$}}}%
  \end{picture}%
\endgroup%

%% file: Images/SeedVertex_new.pdf_tex
\begingroup%
  \makeatletter%
  \providecommand\color[2][]{%
    \errmessage{(Inkscape) Color is used for the text in Inkscape, but the package 'color.sty' is not loaded}%
    \renewcommand\color[2][]{}%
  }%
  \providecommand\transparent[1]{%
    \errmessage{(Inkscape) Transparency is used (non-zero) for the text in Inkscape, but the package 'transparent.sty' is not loaded}%
    \renewcommand\transparent[1]{}%
  }%
  \providecommand\rotatebox[2]{#2}%
  \ifx\svgwidth\undefined%
    \setlength{\unitlength}{154.38399658bp}%
    \ifx\svgscale\undefined%
      \relax%
    \else%
      \setlength{\unitlength}{\unitlength * \real{\svgscale}}%
    \fi%
  \else%
    \setlength{\unitlength}{\svgwidth}%
  \fi%
  \global\let\svgwidth\undefined%
  \global\let\svgscale\undefined%
  \makeatother%
  \begin{picture}(1,1.46981582)%
    \put(0,0){\includegraphics[width=\unitlength]{SeedVertex_new.pdf}}%
    \put(0.44620122,1.37012527){\color[rgb]{0,0,0}\makebox(0,0)[lb]{\smash{$Y$}}}%
    \put(-0.0719872,0.17829173){\color[rgb]{0,0,0}\makebox(0,0)[lb]{\smash{$1$}}}%
    \put(0.13528817,0.0228352){\color[rgb]{0,0,0}\makebox(0,0)[lb]{\smash{$2$}}}%
    \put(1.01620849,0.17829205){\color[rgb]{0,0,0}\makebox(0,0)[lb]{\smash{$k$}}}%
  \end{picture}%
\endgroup%

%% file: Images/ClusterMutation1.pdf_tex
\begingroup%
  \makeatletter%
  \providecommand\color[2][]{%
    \errmessage{(Inkscape) Color is used for the text in Inkscape, but the package 'color.sty' is not loaded}%
    \renewcommand\color[2][]{}%
  }%
  \providecommand\transparent[1]{%
    \errmessage{(Inkscape) Transparency is used (non-zero) for the text in Inkscape, but the package 'transparent.sty' is not loaded}%
    \renewcommand\transparent[1]{}%
  }%
  \providecommand\rotatebox[2]{#2}%
  \ifx\svgwidth\undefined%
    \setlength{\unitlength}{460.9765625bp}%
    \ifx\svgscale\undefined%
      \relax%
    \else%
      \setlength{\unitlength}{\unitlength * \real{\svgscale}}%
    \fi%
  \else%
    \setlength{\unitlength}{\svgwidth}%
  \fi%
  \global\let\svgwidth\undefined%
  \global\let\svgscale\undefined%
  \makeatother%
  \begin{picture}(1,0.50732427)%
    \put(0,0){\includegraphics[width=\unitlength]{ClusterMutation1.pdf}}%
    \put(0.17060636,0.48061467){\color[rgb]{0,0,0}\makebox(0,0)[lb]{\smash{$Y$}}}%
    \put(-0.0026269,0.00611813){\color[rgb]{0,0,0}\makebox(0,0)[lb]{\smash{$A$}}}%
    \put(0.30975341,0.00611813){\color[rgb]{0,0,0}\makebox(0,0)[lb]{\smash{$B$}}}%
    \put(0.46121017,0.24377281){\color[rgb]{0,0,0}\makebox(0,0)[lb]{\smash{$=$}}}%
    \put(0.79536692,0.48061467){\color[rgb]{0,0,0}\makebox(0,0)[lb]{\smash{$Y$}}}%
    \put(0.62213372,0.00611813){\color[rgb]{0,0,0}\makebox(0,0)[lb]{\smash{$B$}}}%
    \put(0.93451402,0.00611813){\color[rgb]{0,0,0}\makebox(0,0)[lb]{\smash{$A$}}}%
  \end{picture}%
\endgroup%

%% file: Images/ClusterMutation2.pdf_tex
\begingroup%
  \makeatletter%
  \providecommand\color[2][]{%
    \errmessage{(Inkscape) Color is used for the text in Inkscape, but the package 'color.sty' is not loaded}%
    \renewcommand\color[2][]{}%
  }%
  \providecommand\transparent[1]{%
    \errmessage{(Inkscape) Transparency is used (non-zero) for the text in Inkscape, but the package 'transparent.sty' is not loaded}%
    \renewcommand\transparent[1]{}%
  }%
  \providecommand\rotatebox[2]{#2}%
  \ifx\svgwidth\undefined%
    \setlength{\unitlength}{765.0078125bp}%
    \ifx\svgscale\undefined%
      \relax%
    \else%
      \setlength{\unitlength}{\unitlength * \real{\svgscale}}%
    \fi%
  \else%
    \setlength{\unitlength}{\svgwidth}%
  \fi%
  \global\let\svgwidth\undefined%
  \global\let\svgscale\undefined%
  \makeatother%
  \begin{picture}(1,0.40782512)%
    \put(0,0){\includegraphics[width=\unitlength]{ClusterMutation2.pdf}}%
    \put(0.18786474,0.39861699){\color[rgb]{0,0,0}\makebox(0,0)[lb]{\smash{$Y$}}}%
    \put(-0.03174086,0.18946879){\color[rgb]{0,0,0}\makebox(0,0)[lb]{\smash{$A$}}}%
    \put(0.20756528,0.00368665){\color[rgb]{0,0,0}\makebox(0,0)[lb]{\smash{$B$}}}%
    \put(0.38655552,0.18946879){\color[rgb]{0,0,0}\makebox(0,0)[lb]{\smash{$C$}}}%
    \put(0.46970622,0.20205582){\color[rgb]{0,0,0}\makebox(0,0)[lb]{\smash{$=$}}}%
    \put(0.77347967,0.39861699){\color[rgb]{0,0,0}\makebox(0,0)[lb]{\smash{$Y$}}}%
    \put(0.55387407,0.18946879){\color[rgb]{0,0,0}\makebox(0,0)[lb]{\smash{$C$}}}%
    \put(0.79318025,0.00368665){\color[rgb]{0,0,0}\makebox(0,0)[lb]{\smash{$A$}}}%
    \put(0.97217045,0.18946879){\color[rgb]{0,0,0}\makebox(0,0)[lb]{\smash{$B$}}}%
  \end{picture}%
\endgroup%

%% file: Images/ClusterMutation3.pdf_tex
\begingroup%
  \makeatletter%
  \providecommand\color[2][]{%
    \errmessage{(Inkscape) Color is used for the text in Inkscape, but the package 'color.sty' is not loaded}%
    \renewcommand\color[2][]{}%
  }%
  \providecommand\transparent[1]{%
    \errmessage{(Inkscape) Transparency is used (non-zero) for the text in Inkscape, but the package 'transparent.sty' is not loaded}%
    \renewcommand\transparent[1]{}%
  }%
  \providecommand\rotatebox[2]{#2}%
  \ifx\svgwidth\undefined%
    \setlength{\unitlength}{861.2265625bp}%
    \ifx\svgscale\undefined%
      \relax%
    \else%
      \setlength{\unitlength}{\unitlength * \real{\svgscale}}%
    \fi%
  \else%
    \setlength{\unitlength}{\svgwidth}%
  \fi%
  \global\let\svgwidth\undefined%
  \global\let\svgscale\undefined%
  \makeatother%
  \begin{picture}(1,0.33657183)%
    \put(0,0){\includegraphics[width=\unitlength]{ClusterMutation3.pdf}}%
    \put(0.21207606,0.32227536){\color[rgb]{0,0,0}\makebox(0,0)[lb]{\smash{$Y$}}}%
    \put(-0.00140606,0.14261091){\color[rgb]{0,0,0}\makebox(0,0)[lb]{\smash{$A$}}}%
    \put(0.1379301,0.00327477){\color[rgb]{0,0,0}\makebox(0,0)[lb]{\smash{$B$}}}%
    \put(0.40731334,0.14261091){\color[rgb]{0,0,0}\makebox(0,0)[lb]{\smash{$D$}}}%
    \put(0.47860885,0.14869441){\color[rgb]{0,0,0}\makebox(0,0)[lb]{\smash{$=$}}}%
    \put(0.29584441,0.00327477){\color[rgb]{0,0,0}\makebox(0,0)[lb]{\smash{$C$}}}%
    \put(0.76942066,0.32227536){\color[rgb]{0,0,0}\makebox(0,0)[lb]{\smash{$Y$}}}%
    \put(0.55593857,0.14261091){\color[rgb]{0,0,0}\makebox(0,0)[lb]{\smash{$D$}}}%
    \put(0.68598565,0.00327477){\color[rgb]{0,0,0}\makebox(0,0)[lb]{\smash{$A$}}}%
    \put(0.96465797,0.14261091){\color[rgb]{0,0,0}\makebox(0,0)[lb]{\smash{$C$}}}%
    \put(0.85318904,0.00327477){\color[rgb]{0,0,0}\makebox(0,0)[lb]{\smash{$B$}}}%
  \end{picture}%
\endgroup%

%% file: Images/n4k2_Bridges.pdf_tex
\begingroup%
  \makeatletter%
  \providecommand\color[2][]{%
    \errmessage{(Inkscape) Color is used for the text in Inkscape, but the package 'color.sty' is not loaded}%
    \renewcommand\color[2][]{}%
  }%
  \providecommand\transparent[1]{%
    \errmessage{(Inkscape) Transparency is used (non-zero) for the text in Inkscape, but the package 'transparent.sty' is not loaded}%
    \renewcommand\transparent[1]{}%
  }%
  \providecommand\rotatebox[2]{#2}%
  \ifx\svgwidth\undefined%
    \setlength{\unitlength}{422.2394043bp}%
    \ifx\svgscale\undefined%
      \relax%
    \else%
      \setlength{\unitlength}{\unitlength * \real{\svgscale}}%
    \fi%
  \else%
    \setlength{\unitlength}{\svgwidth}%
  \fi%
  \global\let\svgwidth\undefined%
  \global\let\svgscale\undefined%
  \makeatother%
  \begin{picture}(1,1.27115745)%
    \put(0,0){\includegraphics[width=\unitlength]{n4k2_Bridges.pdf}}%
    \put(0.17241849,1.24199745){\color[rgb]{0,0,0}\makebox(0,0)[lb]{\smash{$Y$}}}%
    \put(0.02084572,0.00667942){\color[rgb]{0,0,0}\makebox(0,0)[lb]{\smash{$1$}}}%
    \put(0.32399127,0.00667942){\color[rgb]{0,0,0}\makebox(0,0)[lb]{\smash{$2$}}}%
    \put(0.62713683,0.00667942){\color[rgb]{0,0,0}\makebox(0,0)[lb]{\smash{$3$}}}%
    \put(0.93028239,0.00667942){\color[rgb]{0,0,0}\makebox(0,0)[lb]{\smash{$4$}}}%
  \end{picture}%
\endgroup%

%% file: Images/n4k2_OnShellDiagram.pdf_tex
\begingroup%
  \makeatletter%
  \providecommand\color[2][]{%
    \errmessage{(Inkscape) Color is used for the text in Inkscape, but the package 'color.sty' is not loaded}%
    \renewcommand\color[2][]{}%
  }%
  \providecommand\transparent[1]{%
    \errmessage{(Inkscape) Transparency is used (non-zero) for the text in Inkscape, but the package 'transparent.sty' is not loaded}%
    \renewcommand\transparent[1]{}%
  }%
  \providecommand\rotatebox[2]{#2}%
  \ifx\svgwidth\undefined%
    \setlength{\unitlength}{392.22399902bp}%
    \ifx\svgscale\undefined%
      \relax%
    \else%
      \setlength{\unitlength}{\unitlength * \real{\svgscale}}%
    \fi%
  \else%
    \setlength{\unitlength}{\svgwidth}%
  \fi%
  \global\let\svgwidth\undefined%
  \global\let\svgscale\undefined%
  \makeatother%
  \begin{picture}(1,0.79734032)%
    \put(0,0){\includegraphics[width=\unitlength]{n4k2_OnShellDiagram.pdf}}%
    \put(-0.00308736,0.65987902){\color[rgb]{0,0,0}\makebox(0,0)[lb]{\smash{$Y$}}}%
    \put(-0.00308736,0.1091733){\color[rgb]{0,0,0}\makebox(0,0)[lb]{\smash{$1$}}}%
    \put(0.75158344,0.00719076){\color[rgb]{0,0,0}\makebox(0,0)[lb]{\smash{$2$}}}%
    \put(1.01673805,0.39472441){\color[rgb]{0,0,0}\makebox(0,0)[lb]{\smash{$3$}}}%
    \put(0.75158344,0.76186156){\color[rgb]{0,0,0}\makebox(0,0)[lb]{\smash{$4$}}}%
  \end{picture}%
\endgroup%

%% file: Images/n4k2_Bridges_alt.pdf_tex
\begingroup%
  \makeatletter%
  \providecommand\color[2][]{%
    \errmessage{(Inkscape) Color is used for the text in Inkscape, but the package 'color.sty' is not loaded}%
    \renewcommand\color[2][]{}%
  }%
  \providecommand\transparent[1]{%
    \errmessage{(Inkscape) Transparency is used (non-zero) for the text in Inkscape, but the package 'transparent.sty' is not loaded}%
    \renewcommand\transparent[1]{}%
  }%
  \providecommand\rotatebox[2]{#2}%
  \ifx\svgwidth\undefined%
    \setlength{\unitlength}{422.2394043bp}%
    \ifx\svgscale\undefined%
      \relax%
    \else%
      \setlength{\unitlength}{\unitlength * \real{\svgscale}}%
    \fi%
  \else%
    \setlength{\unitlength}{\svgwidth}%
  \fi%
  \global\let\svgwidth\undefined%
  \global\let\svgscale\undefined%
  \makeatother%
  \begin{picture}(1,1.27115745)%
    \put(0,0){\includegraphics[width=\unitlength]{n4k2_Bridges_alt.pdf}}%
    \put(0.17241849,1.24199745){\color[rgb]{0,0,0}\makebox(0,0)[lb]{\smash{$Y$}}}%
    \put(0.02084572,0.00667942){\color[rgb]{0,0,0}\makebox(0,0)[lb]{\smash{$1$}}}%
    \put(0.32399127,0.00667942){\color[rgb]{0,0,0}\makebox(0,0)[lb]{\smash{$2$}}}%
    \put(0.62713683,0.00667942){\color[rgb]{0,0,0}\makebox(0,0)[lb]{\smash{$3$}}}%
    \put(0.93028239,0.00667942){\color[rgb]{0,0,0}\makebox(0,0)[lb]{\smash{$4$}}}%
  \end{picture}%
\endgroup%

%% file: Images/n4k2_OnShellDiagram_alt.pdf_tex
\begingroup%
  \makeatletter%
  \providecommand\color[2][]{%
    \errmessage{(Inkscape) Color is used for the text in Inkscape, but the package 'color.sty' is not loaded}%
    \renewcommand\color[2][]{}%
  }%
  \providecommand\transparent[1]{%
    \errmessage{(Inkscape) Transparency is used (non-zero) for the text in Inkscape, but the package 'transparent.sty' is not loaded}%
    \renewcommand\transparent[1]{}%
  }%
  \providecommand\rotatebox[2]{#2}%
  \ifx\svgwidth\undefined%
    \setlength{\unitlength}{392.22399902bp}%
    \ifx\svgscale\undefined%
      \relax%
    \else%
      \setlength{\unitlength}{\unitlength * \real{\svgscale}}%
    \fi%
  \else%
    \setlength{\unitlength}{\svgwidth}%
  \fi%
  \global\let\svgwidth\undefined%
  \global\let\svgscale\undefined%
  \makeatother%
  \begin{picture}(1,0.79734032)%
    \put(0,0){\includegraphics[width=\unitlength]{n4k2_OnShellDiagram_alt.pdf}}%
    \put(-0.00308736,0.65987902){\color[rgb]{0,0,0}\makebox(0,0)[lb]{\smash{$4$}}}%
    \put(-0.02348387,0.12956962){\color[rgb]{0,0,0}\makebox(0,0)[lb]{\smash{$Y$}}}%
    \put(0.75158344,0.00719076){\color[rgb]{0,0,0}\makebox(0,0)[lb]{\smash{$1$}}}%
    \put(1.01673805,0.39472441){\color[rgb]{0,0,0}\makebox(0,0)[lb]{\smash{$2$}}}%
    \put(0.75158344,0.76186156){\color[rgb]{0,0,0}\makebox(0,0)[lb]{\smash{$3$}}}%
  \end{picture}%
\endgroup%

%% file: Images/Intertwining.pdf_tex
\begingroup%
  \makeatletter%
  \providecommand\color[2][]{%
    \errmessage{(Inkscape) Color is used for the text in Inkscape, but the package 'color.sty' is not loaded}%
    \renewcommand\color[2][]{}%
  }%
  \providecommand\transparent[1]{%
    \errmessage{(Inkscape) Transparency is used (non-zero) for the text in Inkscape, but the package 'transparent.sty' is not loaded}%
    \renewcommand\transparent[1]{}%
  }%
  \providecommand\rotatebox[2]{#2}%
  \ifx\svgwidth\undefined%
    \setlength{\unitlength}{534.17597656bp}%
    \ifx\svgscale\undefined%
      \relax%
    \else%
      \setlength{\unitlength}{\unitlength * \real{\svgscale}}%
    \fi%
  \else%
    \setlength{\unitlength}{\svgwidth}%
  \fi%
  \global\let\svgwidth\undefined%
  \global\let\svgscale\undefined%
  \makeatother%
  \begin{picture}(1,0.35272272)%
    \put(0,0){\includegraphics[width=\unitlength]{Intertwining.pdf}}%
    \put(0.49571679,0.1559499){\color[rgb]{0,0,0}\makebox(0,0)[lb]{\smash{$=$}}}%
    \put(0.06140299,0.00618652){\color[rgb]{0,0,0}\makebox(0,0)[lb]{\smash{$v_i$}}}%
    \put(0.30102439,0.00618652){\color[rgb]{0,0,0}\makebox(0,0)[lb]{\smash{$v_j$}}}%
    \put(0.66045651,0.33566596){\color[rgb]{0,0,0}\makebox(0,0)[lb]{\smash{$v_j$}}}%
    \put(0.90007792,0.33566596){\color[rgb]{0,0,0}\makebox(0,0)[lb]{\smash{$v_i$}}}%
    \put(1.00491228,0.26078426){\color[rgb]{0,0,0}\makebox(0,0)[lb]{\smash{$u$}}}%
    \put(0.40585876,0.08106821){\color[rgb]{0,0,0}\makebox(0,0)[lb]{\smash{$u$}}}%
  \end{picture}%
\endgroup%

%% file: Images/ClusterMutation_general.pdf_tex
\begingroup%
  \makeatletter%
  \providecommand\color[2][]{%
    \errmessage{(Inkscape) Color is used for the text in Inkscape, but the package 'color.sty' is not loaded}%
    \renewcommand\color[2][]{}%
  }%
  \providecommand\transparent[1]{%
    \errmessage{(Inkscape) Transparency is used (non-zero) for the text in Inkscape, but the package 'transparent.sty' is not loaded}%
    \renewcommand\transparent[1]{}%
  }%
  \providecommand\rotatebox[2]{#2}%
  \ifx\svgwidth\undefined%
    \setlength{\unitlength}{618.7890625bp}%
    \ifx\svgscale\undefined%
      \relax%
    \else%
      \setlength{\unitlength}{\unitlength * \real{\svgscale}}%
    \fi%
  \else%
    \setlength{\unitlength}{\svgwidth}%
  \fi%
  \global\let\svgwidth\undefined%
  \global\let\svgscale\undefined%
  \makeatother%
  \begin{picture}(1,0.55452307)%
    \put(0,0){\includegraphics[width=\unitlength]{ClusterMutation_general.pdf}}%
    \put(0.47639669,0.53462534){\color[rgb]{0,0,0}\makebox(0,0)[lb]{\smash{$Y$}}}%
    \put(0.02390001,0.39241209){\color[rgb]{0,0,0}\makebox(0,0)[lb]{\smash{$1$}}}%
    \put(0.07561392,0.1273783){\color[rgb]{0,0,0}\makebox(0,0)[lb]{\smash{$2$}}}%
    \put(0.9159649,0.37948362){\color[rgb]{0,0,0}\makebox(0,0)[lb]{\smash{$k+1$}}}%
    \put(0.38589736,0.00455777){\color[rgb]{0,0,0}\makebox(0,0)[lb]{\smash{$3$}}}%
    \put(0.09500663,0.38594786){\color[rgb]{0,0,0}\makebox(0,0)[lb]{\smash{\scalebox{0.8}{$\tilde\alpha_1$}}}}%
    \put(0.23075564,0.25019883){\color[rgb]{0,0,0}\makebox(0,0)[lb]{\smash{\scalebox{0.8}{$\tilde\alpha_2$}}}}%
    \put(0.74143047,0.28898426){\color[rgb]{0,0,0}\makebox(0,0)[lb]{\smash{\scalebox{0.8}{$\tilde\alpha_k$}}}}%
    \put(0.83839404,0.38594786){\color[rgb]{0,0,0}\makebox(0,0)[lb]{\smash{\scalebox{0.8}{$\alpha_k$}}}}%
    \put(0.59275299,0.20494916){\color[rgb]{0,0,0}\makebox(0,0)[lb]{\smash{\scalebox{0.8}{$\alpha_3$}}}}%
    \put(0.36650464,0.20494916){\color[rgb]{0,0,0}\makebox(0,0)[lb]{\smash{\scalebox{0.8}{$\alpha_2$}}}}%
    \put(0.17904173,0.30191274){\color[rgb]{0,0,0}\makebox(0,0)[lb]{\smash{\scalebox{0.8}{$\alpha_1$}}}}%
    \put(0.43761126,0.19202068){\color[rgb]{0,0,0}\makebox(0,0)[lb]{\smash{\scalebox{0.8}{$\tilde\alpha_3$}}}}%
  \end{picture}%
\endgroup%